\documentclass[]{article}
\usepackage{graphicx}
\usepackage{epstopdf}
\usepackage{cite}
\usepackage{amssymb}
\usepackage{amsthm}
\usepackage{mathtools}
\usepackage[font=small]{caption}
\usepackage{subcaption}
\usepackage{amsmath}
\usepackage{amsfonts}
\usepackage{tikz}
\usepackage{pgfplots}
\usepackage{float}
\usepackage{relsize}
\usepackage[hmarginratio=2:2]{geometry}
\usetikzlibrary{decorations.pathmorphing}
\usepackage{pifont}
\usepackage[titletoc,title]{appendix} 
\usepackage{authblk}
\usepackage[export]{adjustbox}
\usepackage{chngcntr}
\counterwithin{figure}{section}
\title{Nonlinear $q$-Stokes phenomena for $q$-Painlev\'{e} I}
\author{N. Joshi$^1$, C. J. Lustri$^2$ and S. Luu$^1$\footnote{Electronic address: S.Luu@maths.usyd.edu.au; Corresponding author}}
\date{}
\affil{\textit{School of Mathematics and Statistics, The University of Sydney, New South Wales 2006, Australia}$^1$ \\ \textit{Department of Mathematics, Macquarie University, New South Wales 2109, Australia}$^2$}

\begin{document}

\maketitle

\begin{abstract}
We consider the asymptotic behaviour of solutions of the first $q$-difference Painlev\'{e} equation in the limits $|q|\rightarrow 1$ and $n\rightarrow\infty$. Using asymptotic power series, we describe four families of solutions that contain free parameters hidden beyond-all-orders. These asymptotic solutions exhibit Stokes phenomena, which is typically invisible to classical power series methods. In order to investigate such phenomena we apply exponential asymptotic techniques to obtain mathematical descriptions of the rapid switching behaviour associated with Stokes curves. Through this analysis, we also determine the regions of the complex plane in which the asymptotic behaviour is described by a power series expression, and find that the Stokes curves are described by curves known as $q$-spirals.
\end{abstract}

\section{Introduction}\label{S:Introduction}
In this paper we study the first $q$-difference Painlev\'{e} equation ($\text{$q$-P}_\text{I}$) 
\begin{equation}\label{qPainleveI eqn}
\overline{w}\underline{w}=\frac{1}{w}-\frac{1}{xw^2},
\end{equation}
where $w=w(x), \overline{w}=w(qx), \underline{w}=w(x/q)$, and $q\in\mathbb{C}$ such that $|q|\neq 0,1$. In particular, we assume that $|q|>1$ and consider \eqref{qPainleveI eqn} under the limits $|q|\rightarrow 1$ and $n\rightarrow\infty$. Equation \eqref{qPainleveI eqn} is part of a class of integrable, second-order nonlinear difference equations known as the discrete Painlev\'{e} equations that tend to the ordinary Painlev\'{e} equations in the continuum limit. More generally, equation \eqref{qPainleveI eqn} is known as a $q$-difference equation since the evolution of the independent variable $x$ takes the form $x=x_0q^n$ for some initial $x_0$. 

Motivated by Boutroux's study of the first Painlev\'{e} equation \cite{Boutroux1913}, the asymptotic behaviour of \eqref{qPainleveI eqn} in the limit $|x|\rightarrow\infty$ has been considered in \cite{Joshi2015,Joshi2016}. Joshi \cite{Joshi2015} showed that there exists a true solution satisfying $w\rightarrow 0$ as $|x|\rightarrow \infty$, which is asymptotic to a divergent series. However, \cite{Joshi2015} does not describe the Stokes switching behaviour that is typically associated with (divergent) asymptotic series.

In this study we uncover the Stokes behaviour present in the solutions of \eqref{qPainleveI eqn} using exponential asymptotic methods. We first introduce a parameter, $\epsilon$, such that the double limit $|q|\rightarrow 1$ and $n\rightarrow\infty$ is equivalent to $\epsilon\rightarrow 0$. Since we may parametrize $x$ by $x=x_0q^n$, one possible scaling choice which captures the desired behaviour is $n=s/\epsilon$ and $q=1+\epsilon$. Under this choice of scaling we have $x\sim x_0e^s+\mathcal{O}(\epsilon)$ as $\epsilon\rightarrow 0$. We note that the large $x$ limit may be obtained if $s$ is taken to be sufficiently large and positive.  

Exponential asymptotic techniques for differential-difference equations were developed by King and Chapman \cite{King2001} in order to study a nonlinear model of atomic lattices based on the works of \cite{Olde1995,Chapman1998}. The authors of \cite{Chris2015,Luu2017} applied the Stokes smoothing technique described in \cite{King2001} to the first and second discrete Painlev\'{e} equations and obtained asymptotic approximations which contain exponentially small contributions. The difference equations considered in \cite{Chris2015,Luu2017} are of additive type as the independent variable is of the form $z_n=\alpha n+\beta$. Motivated by their work, we extend this to $\text{$q$-P}_\text{I}$ in order to study asymptotic solutions of $q$-difference equations which display Stokes phenomena.

We note that there are other exponential asymptotic approaches used to study difference equations \cite{Olde2004,Olver2000,Immink1988}. In particular, Olde Daalhuis \cite{Olde2004} considered a certain class of second-order linear difference equations, and applied Borel summation techniques in order to obtain asymptotic expansions with exponentially small error. The methods used in these studies involves applying Borel resummation to the divergent asymptotic series of the problem, which allows the optimally-truncated error to be computed directly. We note that Borel summation techniques may be applied whenever a series possess factorial-over-power divergence in the late-order terms and therefore could be applied as an alternative approach to that utilised in the present study.

\subsection{Background}\label{S:Background}
General solutions to the Painlev\'{e} equations are higher transcendental functions, which cannot be expressed exactly in terms of elementary functions and therefore much of the analytic information regarding their general solutions are very limited. In fact, Nishioka \cite{Nishioka2010} proved that the general solutions of $\text{$q$-P}_\text{I}$ are not expressible in terms of solutions of first-order $q$-difference equations. 

The discrete Painlev\'{e} equations have appeared in areas of physical interest such as in mathematical physics models describing two-dimensional quantum gravity \cite{Brezin1990,Periwal1990matrixmodelsolve,Periwal1990,Forrester2015,Fokas1991,Vera1996}. The basis of these models have origins in orthogonal polynomial theory, in which discrete Painlev\'{e} equations have been found to commonly appear \cite{Shohat1939,Laguerre1885,Magnus1995,Magnus1999,VanAssche2006,VanAssche2015,Knizel2015}. Since the discrete Painlev\'{e} equations often arise in many nonlinear models of mathematical physics they are often regarded as defining new nonlinear special functions \cite{Clarkson2003,Its1992}. 

Motivated by these applications, much research has gone into the asymptotic study of the (discrete) Painlev\'{e} equations. Previous asymptotics studies for the first discrete Painlev\'{e} equation have been conducted in \cite{Joshi1997,Veresh1995,Chris2015} where the authors found solutions asymptotically free of poles in the large independent variable limit, which share features with the (tri)-tronque\'{e} solutions of the first Painlev\'{e} equation found by Boutroux \cite{Boutroux1913}. Asymptotic solutions have also been found for the so called alternate discrete Painlev\'{e} I equation \cite{Joshi2017,Clarkson2016withAssche}.

Extending the work of \cite{Chris2015}, the authors of \cite{Luu2017} also find solutions of the second discrete Painlev\'{e} equation ($\text{dP}_{\text{II}}$), which are asymptotically free of poles in some domain of the complex plane containing the positive real axis. This was achieved by rescaling the problem such that the step size of the rescaled difference equation is small in the asymptotic limit. A similar study on $\text{dP}_{\text{II}}$ was also investigated by Shimomura \cite{Shinomura2012}. Shimomura showed that an asymptotic solution of $\text{dP}_{\text{II}}$ which reduces to the tri-tronque\'{e} solution of $\text{P}_\text{II}$ can be found by choosing a scaling of $\text{dP}_{\text{II}}$ such that the rescaled equation tends to the second continuous Painlev\'{e} equation ($\text{P}_\text{II}$) in the limit $n\rightarrow\infty$.

The isomonodromic deformation method has also been used to study the asymptotics for nonlinear difference equations \cite{Vera1996,Krichever2004,Fokas1992,Cassatella2012}. Using the nonlinear steepest descent method developed by Deift and Zhou \cite{Deift1992}, the authors of \cite{Xu2013} find the asymptotic behaviour of solutions of the fifth discrete Painlev\'{e} equation in terms of the solutions of the fifth continuous Painlev\'{e} equation. 

However, there have been very few asymptotic studies on $q$-Painlev\'{e} equations. The asymptotic study of variants of the sixth $q$-Painlev\'{e} equation have been investigated by Mano \cite{Mano2010} and Joshi and Roffelson \cite{Roffelson2016} using the $q$-analogue of the isomonodromic deformation approach. The first $q$-Painlev\'{e} equation has also been investigated in \cite{Joshi2015,Joshi2016}. In particular, Joshi \cite{Joshi2015} proves the existence of true solutions of \eqref{qPainleveI eqn}, which are asymptotic to a divergent asymptotic power series in the limit $|x|\rightarrow \infty$. Such expansions are known to exhibit Stokes phenomena in the complex plane. 

Stokes phenomena are generally well understood in the case of linear and nonlinear differential equations \cite{Paris1992part1,Braaksma2002,Howls2012,Chapman2005,Chapman2007}. The study of Stokes phenomena have also been investigated for nonlinear difference equations \cite{King2001}, and have been extended to study discrete Painlev\'{e} equations \cite{Chris2015,Luu2017}. 

In the continuous theory, Borel summation methods are also used to describe Stokes behaviour by resumming divergent asymptotic series expansions. The $q$-analogues of these methods have also been developed for linear $q$-difference equations \cite{Zhang1998,Zhang2002,Zhang2006,DiVizio2009,Sauloy2013} in order to describe behaviour known as $q$-Stokes phenomena. These methods have been explicitly applied to certain classes of second-order linear $q$-difference equations by Morita \cite{Morita2011,Morita2014,Morita2014part2} and Ohyama \cite{Ohyama2016}.

However, $q$-Stokes phenomenon is unlike classical Stokes phenomenon for differential equations. The notion of $q$-Stokes phenomenon and its differences to classical Stokes phenomenon are detailed in \cite{DiVizio2009,Sauloy2013}. To the best of our knowledge there have been no corresponding studies for nonlinear $q$-difference equations. The goal of this study is to extend the exponential asymptotic methods used in \cite{King2001,Chris2015,Luu2017} to $q$-difference equations in order to describe Stokes behaviour present in the asymptotic series expansions.

\subsection{Exponential asymptotics and Stokes curves}\label{S:Exponential asymptotics and Stokes curves}
Conventional asymptotic power series methods fail to capture the presence of exponentially small terms, and therefore these terms are often described as lying beyond-all-orders. In order to investigate such terms, exponential asymptotic methods are used. The underlying principle of these methods is that divergent asymptotic series may be truncated so that the divergent tail, also known as the remainder term, is exponentially small in the asymptotic limit \cite{Boyd1999}. This is known as an optimally-truncated asymptotic series. Thereafter, the problem can be rescaled in order to directly study the behaviour of these exponentially small remainder terms. This idea was introduced by Berry \cite{Berry1988,Berry1989,Berry1991}, and Berry and Howls \cite{Berry1990}, who used these methods to determine the behaviour of special functions such as the Airy function. 

The basis of this study uses techniques of exponential asymptotics developed by Olde Daalhuis et al. \cite{Olde1995} for linear differential equations, extended by Chapman et al. \cite{Chapman1998} for application to nonlinear differential equations, and further developed by King and Chapman \cite{King2001} for nonlinear differential-difference equations. A brief outline of the key steps of the process will be provided here, however more detailed explanation of the methodology may be found in these studies.

In order to optimally truncate an asymptotic series, the general form of the coefficients of the asymptotic series is needed. However, in many cases this is an algebraically intractable problem. Dingle \cite{Dingle1973} investigated singular perturbation problems and noted that the calculation of successive terms of the asymptotic series involves repeated differentiation of the earlier terms. Hence, the late-order terms, $a_m$, of the asymptotic series typically diverge as the ratio between a factorial and an increasing power of some function as $m\rightarrow \infty$. A typical form describing this is given by the expression 
\begin{equation}\label{general late order terms}
a_m\sim \frac{A\,\Gamma(m+\gamma)}{\chi^{m+\gamma}},
\end{equation}
as $m\rightarrow\infty$ where $\Gamma$ is the gamma function defined in \cite{Abram2012}, while $A$, and $\chi$ are functions of the independent variable which do not depend on $m$, known as the prefactor and singulant respectively. The singulant is subject to the condition that it vanishes at the singular points of the leading order behaviour, ensuring that the singularity is present in all higher-order terms. Chapman et al. \cite{Chapman1998} noted this behaviour in their investigations and utilize \eqref{general late order terms} as an ansatz for the late-order terms, which may then be used to optimally truncate the asymptotic expansion.

Following \cite{Olde1995} we substitute the optimally-truncated series back into the governing equation and study the exponentially small remainder term. When investigating these terms we will discover two important curves known as Stokes and anti-Stokes curves \cite{Bender1999}. Stokes curves are curves on which the switching exponential is maximally subdominant compared to the leading order behaviour. As Stokes curves are crossed, the exponentially small behaviour experiences a smooth, rapid change in value in the neighbourhood of the curve; this is known as Stokes switching.  Anti-Stokes curves are curves along which the exponential term switches from being exponentially small to exponentially large (and vice versa). We will use these definitions to determine the locations of the Stokes and anti-Stokes curves in this study.

By studying the switching behaviour of the exponentially small remainder term in the neighbourhood of Stokes curves, it is possible to obtain an expression for the remainder term. The behaviour of the remainder associated with the late-order terms in \eqref{general late order terms} typically takes the form $\mathcal{S}A\exp(-\chi/\epsilon)$, where $\mathcal{S}$ is a Stokes multiplier that is constant away from Stokes curves, but varies rapidly between constant values as Stokes curves are crossed.  From this form, it can be shown that Stokes lines follow curves along which $\chi$ is real and positive, while anti-Stokes lines follow curves along which $\chi$ is imaginary. A more detailed discussion of the behaviour of Stokes curves is given in \cite{Bender1999}. 

\subsection{Paper outline}\label{S:Paper outline}
In Section \ref{S:Asymptotic series expansions}, we find two classes of solution behaviour of $\text{$q$-P}_\text{I}$, which we refer as Type A and Type B solutions. We also determine the formal series expansions of these solutions, and provide the recurrence relations for the coefficients. 

In Section \ref{S:Exponential Asymptotics}, we determine the form of the late-order terms for Type A solutions and use this to determine the Stokes structure of these asymptotic series expansions. We then calculate the behaviours of the exponentially small contributions present in these solutions as Stokes curves are crossed. This is then used to determine the regions in which the asymptotic power series are accurate representations of the dominant asymptotic behaviour. 

In Section \ref{S:Type B Asymptotics}, we consider Type B solutions of $\text{$q$-P}_\text{I}$ following the analysis in Section \ref{S:Exponential Asymptotics}. In Section \ref{S: connection to the nonzero asym behaviours of qpI} we establish a connection between Type A and B solutions to the vanishing and non-vanishing asymptotic solutions of $\text{$q$-P}_\text{I}$ found by Joshi \cite{Joshi2015}. Finally, we discuss the results and conclusions of the paper in Section \ref{S:Conclusions}.  Appendices A-D contain detailed calculations needed in Section \ref{S:Exponential Asymptotics}.

\section{Asymptotic series expansions}\label{S:Asymptotic series expansions}
In this section, we expand the solution as a formal power series in the limit $\epsilon\rightarrow 0$, obtain the recurrence relation for the coefficients of the series and deduce the general expression of the late-order terms.

We first rewrite \eqref{qPainleveI eqn} as an additive difference equation by setting $x=x_0q^n$, which gives
\begin{equation}\label{qPI discrete time}
w_{n+1}w_{n-1}=\frac{1}{w_n}-\frac{1}{x_0q^nw_{n}^2},
\end{equation} 
where $w_n=w(x_0q^n)$. In our analysis, we will introduce a small parameter, $\epsilon$, by rescaling the variables appearing in \eqref{qPI discrete time}. The choice of scalings we apply are given by 
\begin{equation}\label{qPI scalings}
s=\epsilon n, \qquad q=1+\epsilon, \qquad w_n=W(s).
\end{equation}
Under these scalings, equation \eqref{qPI discrete time} becomes
\begin{equation}\label{qPI rescaled}
W(s+\epsilon)W(s)^2W(s-\epsilon)=W(s)-\frac{1}{x_0(1+\epsilon)^{s/\epsilon}},
\end{equation} 
and consider the limit $\epsilon\rightarrow 0$. We also note that under the scalings given by \eqref{qPI scalings}, the independent variable, $x$, has behaviour described by
\begin{equation}\label{q difference rescale independent variable}
x = (1+\epsilon)^{s/\epsilon} \sim x_0e^s +\mathcal{O}(\epsilon),
\end{equation}
as $\epsilon\rightarrow 0$. As $e^s$ is an entire function, we set $x_0=1$ for the remainder of this analysis. It can be shown that equation \eqref{qPI rescaled} is invariant under the mapping 
\begin{equation}\label{rescaled qPI symmetry}
W\mapsto \lambda W, \quad \text{with} \quad s\mapsto s-\frac{\epsilon\log(\lambda)}{\log(1+\epsilon)} \sim s-\log(\lambda)+\mathcal{O}(\epsilon),
\end{equation}
as $\epsilon\rightarrow 0$, and where $\lambda^3=1$. We note that this corresponds to the rotational symmetry of \eqref{qPainleveI eqn} found in \cite{Joshi2015}. 

We expand the solution, $W(s)$, as an asymptotic power series in $\epsilon$ by writing
\begin{equation}\label{qPI asymptotic power series}
W(s) \sim \sum_{r=0}^{\infty}\epsilon^{r}W_r(s),
\end{equation}
as $\epsilon\rightarrow 0$. Substituting \eqref{qPI asymptotic power series} into \eqref{qPI rescaled} and matching terms of $\mathcal{O}(\epsilon^r)$ we obtain the nonlinear recurrence relation
\begin{equation}\label{even recurrence relation}
\sum_{q=0}^{r}\sum_{m=0}^{q}\sum_{k=0}^{m}\frac{(-1)^kW_{m-k}^{(k)}}{k!}\sum_{j=0}^{q-m}\frac{W_{q-m-j}^{(j)}}{j!}\sum_{b=0}^{r-q}W_{r-q-b}W_b=W_r-e^{-s}P_r(-s)
\end{equation}
for $r\geq 0$ and where the polynomials $P_n(s)$ are given by  
\begin{equation*}\label{intext euler series coeffs}
P_n(s)=\sum_{r=0}^{n}s^r\sum_{k=0}^{r}\frac{(-1)^{r-k}}{(r-k)!}\frac{s_1(k+n,k)}{(k+n)!},
\end{equation*}
where $s_1(n,k)$ are the Stirling numbers of the first kind \cite{Abram2012}. This recurrence relations allows us to calculate $W_r$ in terms of the previous coefficients. From \eqref{even recurrence relation} we find that the leading order behaviour satisfies
\begin{equation}\label{qPI leading order equation} 
W_0^4 = W_0-e^{-s}.
\end{equation}
Equation \eqref{qPI leading order equation} is invariant under $s\mapsto s+2\pi i$ as the function $e^s$ is $2\pi i$-periodic and hence $W_{0}(s)$ is $2\pi i$-periodic. Furthermore, it will be shown in Section \ref{S:Stokes smoothing} that the Stokes switching behaviour of \eqref{qPI asymptotic power series} depends on $W_{0}(s)$ to leading order. Hence we restrict our attention the domain, $\mathcal{D}_0$, described by
\begin{equation}\label{base sym domain}
\mathcal{D}_0 = \left\{s\in\mathbb{C} \ | \ \text{Im}(s)\in (-\pi,\pi]\right\}.
\end{equation}
We also define the domain, $\mathcal{D}_k$, which we call $k^{\text{th}}$-adjacent domain, by 
\begin{equation}\label{kth adjacent domain}
\mathcal{D}_k = \left\{s\in\mathbb{C} \ | \ \text{Im}(s)\in (-\pi+2k\pi,\pi+2k\pi]\right\}.
\end{equation}
As $W_0$ satisfies a quartic we therefore have four possible leading order behaviours as $\epsilon\rightarrow 0$. We first define the following
\begin{equation}\label{qPI leading order component A-C}
A = 4\left(\frac{2}{3}\right)^{1/3}e^{-s}, \qquad
B = 9+\sqrt{3}\sqrt{27-256e^{-3s}}, \qquad
C = 2^{1/3}3^{2/3},
\end{equation}
and
\begin{equation}\label{qPI leading order component D}
D = \frac{A}{B^{1/3}}+\frac{B^{1/3}}{C}. 
\end{equation}
Then the four solutions for $W_0$ are given by
\begin{equation}\label{qPI W01-W02}
W_{0,1} = -\frac{\sqrt{D}}{2}+\frac{i}{2}\sqrt{D+\frac{2}{\sqrt{D}}}, \qquad
W_{0,2} = -\frac{\sqrt{D}}{2}-\frac{i}{2}\sqrt{D+\frac{2}{\sqrt{D}}},  
\end{equation}
and
\begin{equation}\label{qPI W03-W04}
W_{0,3} = \frac{\sqrt{D}}{2}+\frac{1}{2}\sqrt{-D+\frac{2}{\sqrt{D}}}, \qquad
W_{0,4} = \frac{\sqrt{D}}{2}-\frac{1}{2}\sqrt{-D+\frac{2}{\sqrt{D}}}.
\end{equation} 
From \eqref{qPI W01-W02} and \eqref{qPI W03-W04} it can be shown that
\begin{equation}\label{qPI w04 expression in terms of others}
W_{0,4}(s)= -\sum_{j=1}^{3}W_{0,j}(s).
\end{equation}
Each of the leading order solutions, $W_0$, are singular at points for which the argument of the square root term of $B$ is equal to zero. The singularities of $W_0$ are given by
\begin{equation}\label{qPI leading order singularities}
s_0 = \frac{1}{3}\left(\log\left(\frac{256}{27}\right)+2ik\pi\right),
\end{equation}
where $k\in\mathbb{Z}$. Let us denote the singularities in $\mathcal{D}_0$ by
\begin{equation}\label{type A singularities}
s_{0,1} = \frac{1}{3}\left(\log\left(\frac{256}{27}\right)-2i\pi\right), \qquad
s_{0,2} = \frac{1}{3}\left(\log\left(\frac{256}{27}\right)+2i\pi\right), \qquad
s_{0,3} = \frac{1}{3}\log\left(\frac{256}{27}\right).
\end{equation}
Then it can be shown that the local behaviour of $W_{0,j}(s)$ near the singular points \eqref{qPI leading order singularities} is given by 
\begin{equation}\label{local qPI leading order behaviour}
W_{0,j}\sim \left(\frac{1}{4}\right)^{1/3}e^{2ij\pi/3}+\left(\frac{1}{8\sqrt{2}}\right)^{1/3}e^{2ij\pi/3}\sqrt{s-s_{0,j}}+\mathcal{O}(s-s_{0,j}),
\end{equation}
as $s\rightarrow s_{0,j}$ for $j=1,2,3$. Equation \eqref{qPI w04 expression in terms of others} shows that $W_{0,4}(s)$ is the sum of $W_{0,j}(s)$ for $j=1,2,3$, and is therefore singular at the points $s_{0,1}, s_{0,2}$ and $s_{0,3}$ in $\mathcal{D}_0$. 

Two types of leading order behaviours can be characterized by the number of points at which they are singular in $\mathcal{D}_0$. In the subsequent analysis, we will refer to solutions with leading order behaviour described by $W_{0,j}$ for $j=1,2,3$ as Type A, while those with leading order behaviour described $W_{0,4}$ as Type B. 

In Sections \ref{S:Stokes structure} and \ref{S:Stokes structure type B} we will show that the (anti-) Stokes curves emanate from these singularities. Consequently, the Stokes structure of Type A solutions will be shown to emerge from a single singularity in $\mathcal{D}_0$, while Type B solutions will have (anti-) Stokes curves emanating from three singular points in $\mathcal{D}_0$. In this sense, Type B solutions will have the most complicated Stokes behaviour due to possible interaction effects between the distinct singularities. 

\section{Type A Exponential Asymptotics}\label{S:Exponential Asymptotics}
In this section we will investigate the Stokes phenomena exhibited in Type A solutions with leading order behaviour $W_{0,3}$ as $\epsilon\rightarrow 0$. This is done by first determining the leading order behaviour of the late-order terms, which will allow us to optimally truncate \eqref{qPI asymptotic power series} and study the optimally-truncated error. The results for the remaining Type A solutions may be obtained using the symmetry \eqref{rescaled qPI symmetry}.

\subsection{Late-order terms}\label{S:Late-order terms}
As discussed in Section \ref{S:Exponential asymptotics and Stokes curves}, the ansatz for the late-order terms is given by a factorial-over-power form since the determination of $W_r$ in \eqref{even recurrence relation} involves repeated differentiation. We therefore assume that the coefficients of \eqref{qPI asymptotic power series} with $W_0=W_{0,3}$ are described by
\begin{equation}\label{even late order terms}
W_r(s)\sim \frac{U_3(s)\Gamma(r+\gamma_1)}{\chi_3(s)^{r+\gamma_1}},
\end{equation}
as $r\rightarrow\infty$, where $\chi_3(s)$ is the singulant, $U_3(s)$ is the prefactor and $\gamma_1$ a constant. We substitute \eqref{even late order terms} into \eqref{even recurrence relation} to obtain 
\begin{align}
&2W_{0,3}^3W_r+W_{0,3}^3\sum_{k=0}^{r}\frac{\left((-1)^k+1\right)}{k!}(-\chi_3')^kW_r+6W_{0,3}^2W_1W_{r-1} \nonumber \\
&+W_{0,3}^3\sum_{k=0}^{r}\frac{\left((-1)^k+1\right)}{k!}\left(\binom{k}{1}(-\chi_3')^{k-1}\frac{U_3'}{U_3}+\binom{k}{2}(-\chi_3')^{k-2}(-\chi_3'')\right)W_{r-1} \nonumber \\
&+\sum_{k=0}^{r-1}\frac{\left((-1)^k+1\right)}{k!}(-\chi_3')^k\left(3W_{0,3}^2W_1+W_{0,3}^2W_{0,3}'\right)W_{r-1}+\mathcal{O}(W_{r-2})=W_r+\cdots, \label{even late order terms with sing and prefac} 
\end{align}
where the remaining terms are negligible for the purposes of this demonstration as $r\rightarrow\infty$. By matching terms of $\mathcal{O}(W_r)$ as $r\rightarrow\infty$, we find that the leading order equation is given by
\begin{equation}\label{even singulant equation}
2W_{0,3}^3+W_{0,3}^2\sum_{k=0}^{r}\frac{\left((-1)^k+1\right)}{k!}(-\chi_3')^k=1,
\end{equation}
as $r\rightarrow\infty$. Matching at the next subsequent order involves matching terms of $\mathcal{O}(W_{r-1})$ in \eqref{even late order terms with sing and prefac}. Doing so gives
\begin{align}\label{even prefactor equation}
W_{0,3}^3&\sum_{k=0}^{r}\frac{\left((-1)^k+1\right)}{k!}\left(\binom{k}{1}(-\chi_3')^{k-1}\frac{U_3'}{U_3}+\binom{k}{2}(-\chi_3')^{k-2}(-\chi_3'')\right)+6W_{0,3}^2W_1 \nonumber \\
&+3W_{0,3}^2W_1\sum_{k=0}^{r-1}\frac{\left((-1)^k+1\right)}{k!}(-\chi')^{k}+W_{0,3}^2W_{0,3}'\sum_{k=0}^{r-1}\frac{\left((-1)^k-1\right)}{k!}(-\chi_3')^{k}=0.
\end{align}
as $r\rightarrow\infty$. In order to determine the singulant, $\chi_3(s)$, we consider \eqref{even singulant equation}. The leading order behaviour of $\chi_3(s)$ may be determined by replacing the upper limit of the sum in \eqref{even singulant equation} by infinity. Taking the series to be infinite introduces error in the singulant behaviour which is exponentially small in the limit $r\rightarrow\infty$, which is negligible here \cite{King2001,Chris2015,Luu2017}. Evaluating the sum appearing in \eqref{even singulant equation} as an infinite series gives 
\begin{equation}\label{qPI SINGULANT DE}
\cosh(-\chi_3')=\frac{1-2W_{0,3}^3}{2W_{0,3}^3}, \quad \chi(s_{0,3})=0. 
\end{equation}
The solution of \eqref{qPI SINGULANT DE} is given by 
\begin{equation}\label{qPI singulant expressions}
\chi_3(s;M) = \pm\int_{s_{0,3}}^{s}\left(\cosh(\sigma(t))+2iM\pi\right) \ dt, \qquad \sigma(s)=\frac{1-2W_{0,3}(s)^3}{2W_{0,3}(s)^3}
\end{equation}
where $M\in\mathbb{Z}$. Noting that there are two different expressions for the singulant, we name them $\chi_3(s;M)$ and $\chi_3^{-}(s;M)$ with the choice of the positive and negative signs respectively. In general, the behaviour of $W_r$ will be the sum of expressions \eqref{even late order terms}, with each value of $M$ and sign of the singulant \cite{Dingle1973}. However, this sum will be dominated by the two terms associated with $M=0$ as this is the value for which $|\chi|$ is smallest \cite{Chapman1998}. Thus, we consider the $M=0$ case in the subsequent analysis. Hence we set
\begin{equation}\label{qPI plus singulant}
\chi_3(s)=\int_{s_{0,3}}^{s}\cosh^{-1}(\sigma(t))dt, \qquad \chi_3^{-}(s)=-\int_{s_{0,3}}^{s}\cosh^{-1}(\sigma(t))dt.
\end{equation}
In order to determine the prefactor associated with each singulant we solve \eqref{even prefactor equation}. As before, we replace the upper limit of the sum appearing in \eqref{even prefactor equation} by infinity and use \eqref{qPI singulant expressions} in order to obtain
\begin{equation}\label{qPI PREFACTOR DE}
-2W_{0,3}^3\sinh(\chi')\frac{U_3'}{U_3}-W_{0,3}^3\chi_3''\cosh(\chi_3')+2W_{0,3}^2W_{0,3}'\sinh(\chi_3')+3\frac{W_1}{W_{0,3}}=0.
\end{equation}
It can be verified that 
\begin{equation}\label{prefactor DE ASIDE soln}
F(s)=\frac{\Upsilon W_{0,3}}{\sqrt{\sinh(\chi_3')}},
\end{equation}
where $\Upsilon$ is a constant of integration, solves the differential equation \eqref{qPI PREFACTOR DE} without the $3W_1/W_{0,3}$ term. By setting $U_3(s)=F(s)\phi(s)$ and substituting this into \eqref{qPI PREFACTOR DE} we obtain the differential equation
\begin{equation}\label{prefactor DE part 2 FINAL}
\frac{\phi'}{\phi}=-\frac{W_1\chi_3''}{W_{0,3}'},
\end{equation}
where we have used \eqref{qPI SINGULANT DE} in order to express \eqref{prefactor DE part 2 FINAL} in terms of $\chi_3''$. Integrating \eqref{prefactor DE part 2 FINAL} then gives  
\begin{equation}\label{prefactor second part expression}
\phi(s)=\tilde{\Upsilon}\exp\left(-\int_{}^{s}\frac{W_1(t)\chi_3''(t)}{W_{0,3}'(t)}dt\right),
\end{equation}
where $\tilde{\Upsilon}$ is a constant of integration. Hence, if we let $U_3$ and $U_3^{-}$ denote the prefactors associated with the singulants $\chi_3$ and $\chi_3^{-}$, respectively, then the solutions of \eqref{qPI PREFACTOR DE} are given by
\begin{equation}\label{qPI EXACT PREFACTOR}
U_3(s)=\frac{\Lambda W_{0,3}e^{-G}}{\sqrt{\sinh(\chi_3')}}, \qquad U_3^{-}(s)=\frac{\tilde{\Lambda} W_{0,3}e^{G}}{\sqrt{\sinh(\chi_3')}},
\end{equation}
where $\Lambda, \tilde{\Lambda}$ are constants (which depend on $\Upsilon$ and $\tilde{\Upsilon}$) and 
\begin{equation}\label{integral expression in prefactor}
G(s)= \int_{}^{s}\frac{W_1(t)\chi_3''(t)}{W_{0,3}'(t)}dt.
\end{equation}
Substituting the \eqref{qPI plus singulant} and \eqref{qPI EXACT PREFACTOR} into \eqref{even late order terms} shows that
\begin{equation}\label{late order terms with sing and prefac}
W_r(s)\sim \frac{W_{0,3}\Gamma(r+\gamma_1)}{\sqrt{\sinh(\chi_3')}\chi_3^{r+\gamma_1}}\left(\Lambda e^{-G}+\frac{\tilde{\Lambda}e^G}{(-1)^{r+\gamma_1}}\right),
\end{equation}
as $r\rightarrow\infty$.  

To completely determine the form of the $W_r$ we must also determine the value of $\gamma_1$. This requires matching the late-order expression given in \eqref{even late order terms} to the leading-order behaviour in the neighbourhood of the singularity. This procedure is described in Appendix \ref{S:Appendix calculate prefactors}, and shows that $\gamma_1 = -1/2$ and $\tilde{\Lambda}=-i\Lambda$. 

From the results obtained in Appendix \ref{S:Appendix calculate prefactors}, we find that 
\begin{equation}\label{explicit LOT}
W_{2r}(s)\sim\frac{2W_{0,3}\Lambda\cosh(G)\Gamma(2r-1/2)}{\sqrt{\sinh(\chi_3')}\chi_3^{2r-1/2}}, \qquad
W_{2r+1}(s)\sim-\frac{2W_{0,3}\Lambda\sinh(G)\Gamma(2r+1/2)}{\sqrt{\sinh(\chi_3')}\chi_3^{2r+1/2}},
\end{equation} 
as $r\rightarrow\infty$. Hence, \eqref{explicit LOT} reveals that the asymptotic series \eqref{qPI asymptotic power series} may be expressed as the sum of two asymptotic series in even and odd powers of $\epsilon$.

In the following analysis, we apply optimal truncation methods to the asymptotic series, \eqref{qPI asymptotic power series} and show that the optimal truncation error is proportional to $U_3e^{-\chi_3/\epsilon}$. 

\subsection{Stokes smoothing}\label{S:Stokes smoothing}
In order to determine the behaviour of the exponentially small contributions in the neighbourhood of the Stokes curve we optimally truncate \eqref{qPI asymptotic power series}. We truncate the asymptotic series at the least even term by writing 
\begin{equation}\label{qPI optimally truncated series}
W(s) = \sum_{r=0}^{2N_{\text{opt}}-1}\epsilon^r W_r(s) + R_N(s),
\end{equation}
where $N_{\text{opt}}$ is the optimal truncation point and $R_N$ is the optimally-truncated error. As the analysis is technical, we will summarize the key results in this section with the details provided in Appendix \ref{S:Appendix Stokes smoothing}.

In Appendix \ref{S:Appendix Stokes smoothing}, we show that the optimal truncation point is given by $N_{\text{opt}} \sim (|\chi_3|/\epsilon+\kappa)/2,$ as $\epsilon\rightarrow 0$, and where $\kappa \in [0,1)$ is chosen such that $N_\text{opt}\in\mathbb{Z}$. 

The leading order behaviour of the remainder term in the small $\epsilon$ limit can be shown to take the form
\begin{equation}\label{explicit R_N FORM}
R_N \sim \mathcal{S}_3U_3e^{-\chi_3/\epsilon},
\end{equation}
where $\mathcal{S}_3$ is the Stokes switching parameter which varies rapidly between constant values in the neighbourhood of Stokes curves. 

For algebraic convenience, we let $T(s)$ represent the finite sum in \eqref{qPI optimally truncated series} and use the shift notation denoted by $F=F(s), \overline{F}=F(s+\epsilon), \underline{F}=F(s-\epsilon)$. By substituting the truncated series, \eqref{qPI optimally truncated series}, into the governing equation \eqref{qPI rescaled} we obtain
\begin{equation}\label{governing qPI remainder equation}
T^2\underline{T}\overline{R}+2\overline{T}T\underline{T}R+\overline{T}T^2\underline{R}+\epsilon^{2N}(W_{2N}-4W_{0,3}^3W_{2N})+\cdots = R,
\end{equation}
where the terms neglected are of order $\mathcal{O}(\epsilon^{2N+1}W_{2N+1})$ and quadratic in $R_N$. Terms of these sizes are negligible compared to the terms kept in \eqref{governing qPI remainder equation} as $\epsilon\rightarrow 0$.

Recall that Stokes switching occurs across Stokes curves, which are characterized by $\text{Im}(\chi)=0$ and $\text{Re}(\chi)>0$. We define the value of the Stokes multiplier, $\mathcal{S}_3$, to be
\begin{equation}\label{Stokes multiplier notation defn}
\mathcal{S}_3 = \begin{cases}
\mathcal{S}_3^+ , \quad \text{in the regions where Im$(\chi)_3>0$ and Re$(\chi_3)>0$,}\\
\mathcal{S}_3^- , \quad \text{in the regions where Im$(\chi)_3<0$ and Re$(\chi_3)>0$.}
\end{cases}
\end{equation} 
In Appendix \ref{S:Appendix Stokes smoothing} we show that the Stokes multiplier, $\mathcal{S}_3$, changes in value by 
\begin{equation*}\label{qPI Stokes Multiplier JUMP}
\Delta\mathcal{S}_3=\mathcal{S}_3^+-\mathcal{S}_3^-\sim i\sqrt{2\pi\epsilon}H(\left|\chi_3\right|)\int_{-\infty}^{\infty}e^{-x^2/2}dx=2i\pi\sqrt{\epsilon} H(\left|\chi_3\right|),
\end{equation*}
as Stokes curves are crossed and $H$ is the function defined by $H(\chi_3')=(1-4W_{0,3}^3)^{1/2}/\chi_3'$. Let $\mathcal{S}_j$ denote the Stokes multiplier associated with $\chi_j$ to be
\begin{equation}\label{explicit Stokes mult}
\mathcal{S}_j = i\pi\sqrt{\epsilon} H(\left|\chi_j\right|)\left(\text{erf}\left(\theta\sqrt{\frac{\left|\chi_j\right|}{2\epsilon}}\right)+C_j\right),
\end{equation}
for $j=1,2,3$ and $C_j$ is an arbitrary constant. Then the asymptotic power series expansion of Type A solutions of \eqref{qPI rescaled} with leading order behaviour, $W_{0,3}$, up to exponentially small corrections is given by
\begin{equation}\label{qPI type A complete asym}
W(s)\sim W_{0,3}(s)+\sum_{r=1}^{2N-1}\epsilon^{2r}W_{r}(s)+\mathcal{S}_3(s)U_3(s)e^{-\chi_3(s)/\epsilon},
\end{equation}
as $\epsilon\rightarrow 0$. By using the symmetry given by \eqref{rescaled qPI symmetry} we find that the other Type A solutions are given by
\begin{align}
W(s)&\sim W_{0,1}(s)+\lambda_1\sum_{r=1}^{2N-1}\epsilon^{2r}W_{2r}(s)+\mathcal{S}_1(s)U_1(s)e^{-\chi_1(s)/\epsilon},\label{qPI type A complete asym W01} \\
W(s)&\sim W_{0,2}(s)+\lambda_2\sum_{r=1}^{2N-1}\epsilon^{2r}W_{2r}(s)+\mathcal{S}_2(s)U_2(s)e^{-\chi_2(s)/\epsilon},\label{qPI type A complete asym W02}
\end{align}
as $\epsilon\rightarrow 0$, where $\mathcal{S}_j$ is given by \eqref{explicit Stokes mult}, $\lambda_1=e^{-2i\pi/3}$, $\lambda_2=e^{2i\pi/3}$ and 
\begin{equation*}
\chi_1(s)=\chi_3(s+2i\pi/3), \quad U_1(s)=U_3(s+2i\pi/3), \quad \chi_2(s)=\chi_3(s-2i\pi/3), \quad U_2(s)=U_3(s-2i\pi/3). 
\end{equation*}
In particular, the values of $\Lambda_j$ are 
\begin{equation*}\label{Lambda 1 and 2 values} 
\Lambda_1=\lambda_1\Lambda_3 = e^{-2i\pi/3}\Lambda_3, \qquad \Lambda_2 =\lambda_2\Lambda_3 = e^{-2i\pi/3}\Lambda_3, 
\end{equation*}
where $\Lambda_3$ is given by \eqref{qPI lambda value}. Furthermore, the complete asymptotic series expansions of Type A solutions of \eqref{qPI rescaled} have Stokes multipliers which contain a free parameter in the form of $C_j$. This will be further explained in Section \ref{S:Stokes structure}. 

We have successfully determined a family of asymptotic solutions of \eqref{qPI rescaled} which contains exponentially small error. These exponentially small terms exhibit Stokes switching and therefore the expressions \eqref{qPI type A complete asym}-\eqref{qPI type A complete asym W02} describe the asymptotic behaviour in certain regions of the complex plane. The regions of validity for Type A solutions will be determined in Section \ref{S:Stokes structure}.

\subsection{Stokes structure}\label{S:Stokes structure}
In this section we determine the Stokes structure of Type A solutions in both the complex $s$ and $x$-planes. As demonstrated in Section \ref{S:Stokes smoothing}, we found that the exponential contributions present in the series expansions \eqref{qPI type A complete asym}, \eqref{qPI type A complete asym W01} and \eqref{qPI type A complete asym W02} are proportional to $\exp(-\chi_j/\epsilon)$. These terms are exponentially small when $\text{Re}(\chi_j)>0$ and exponentially large when $\text{Re}(\chi_j)<0$. Hence, by considering \eqref{qPI plus singulant} we may determine the behaviour of these terms and the location of the (anti-) Stokes curves. Once the Stokes structure has been determined in the complex $s$-plane, we may reverse the scalings given by \eqref{qPI scalings} in order to determine the Stokes structure in the complex $x$-plane. 

\begin{figure}[h!]
\centering
\begin{adjustbox}{minipage=\linewidth,scale=1}
  \begin{subfigure}[b]{0.45\textwidth}\centering
  \includegraphics[scale=0.55]{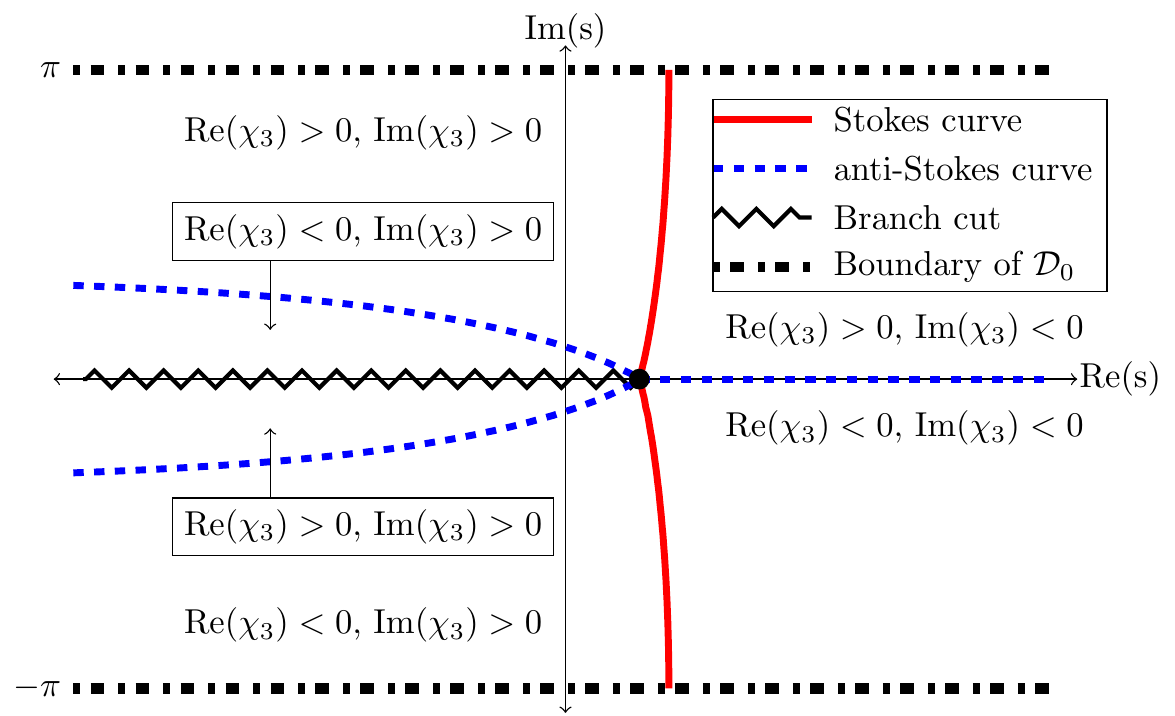}
    \caption{Behaviour of $\chi_3$.}\label{qPI singulant behaviours}
  \end{subfigure}
  \hfill
  \begin{subfigure}[b]{0.45\textwidth}\centering
    \includegraphics[scale=0.55]{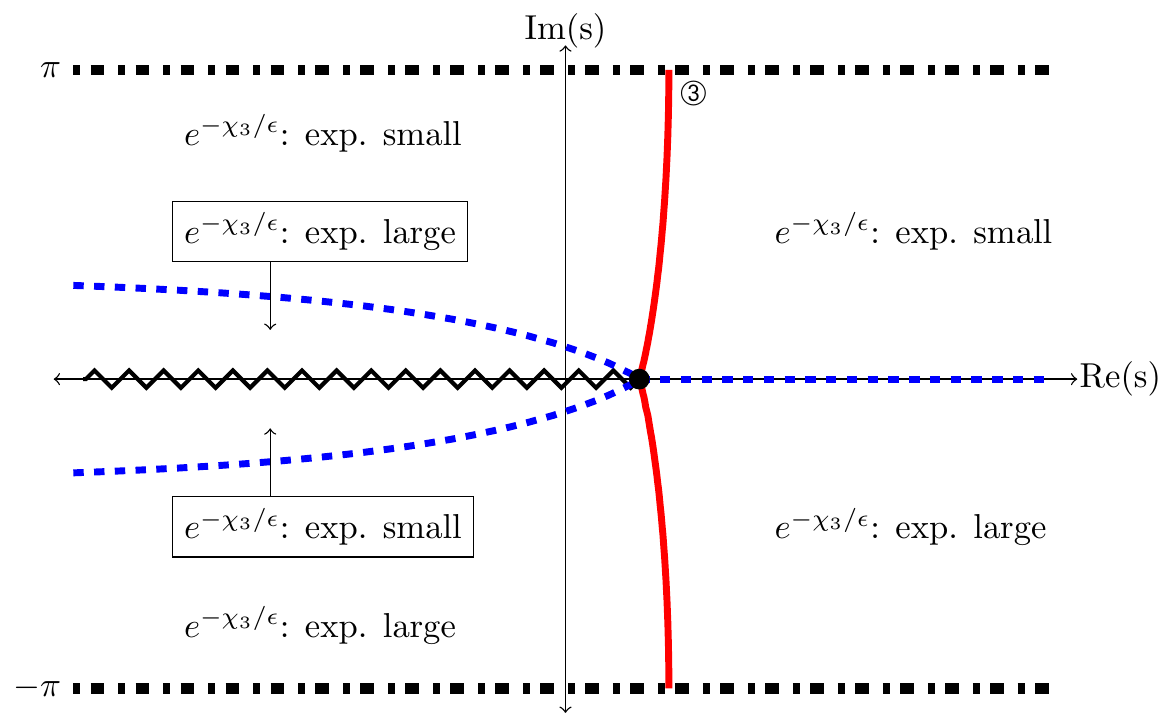}
      \caption{Exponential behaviour.}\label{qPI exponential behaviours}
  \end{subfigure}
    \end{adjustbox}
\caption{These figures illustrate the Stokes structure for the Type A solution described by \eqref{qPI type A complete asym} of \eqref{qPI rescaled} in $\mathcal{D}_0$. Figure \ref{qPI singulant behaviours} illustrates the behaviour of $\chi_3$ as Stokes and anti-Stokes curves are crossed. Figure \ref{qPI exponential behaviours} illustrates regions of $\mathcal{D}_0$ in which the exponential contribution associated with $\chi_3$ is exponentially large or small. The legend of this figure will apply to all subsequent figures unless stated otherwise.}\label{qPI_stokes TOTAL}
\end{figure}

We first illustrate and explain the Stokes structure and Stokes switching behaviour of \eqref{qPI type A complete asym} in the domain $\mathcal{D}_0$, described by \eqref{base sym domain}. The upper and lower boundaries of $\mathcal{D}_0$ are described by the curves $\text{Im}(s)=\pm\pi$ and are denoted by the dot-dashed curves in Figure \ref{qPI singulant behaviours}. In this figure we also see that there are two Stokes curves (red curves) and three anti-Stokes curves (dashed blue curves) emanating from the singularity $s_{0,3}$. The Stokes curves extending towards the upper boundary of $\mathcal{D}_0$ switches the exponential contribution associated with $\chi_3$ as $\text{Re}(\chi_3)>0$. This Stokes curve is denoted by \ding{194} in Figure \ref{qPI exponential behaviours}. While the Stokes curve extending towards the lower boundary of $\mathcal{D}_0$ does not switch any exponential contributions as $\text{Re}(\chi_3)<0$. Additionally, there is a branch cut (zig-zag curve) of $\chi_3$ located along the negative real $s$ axis emanating from $s_{0,3}$. Using this knowledge, we can determine the switching behaviour as Stokes curves are crossed. Additionally, we observe in Figure \ref{qPI singulant behaviours} that the (anti-) Stokes curves and the branch cut separate $\mathcal{D}_0$ into six regions.

\begin{figure}[h!]
\centering
\begin{adjustbox}{minipage=\linewidth,scale=1}
  \begin{subfigure}[b]{0.45\textwidth}\centering
  \includegraphics[scale=0.55]{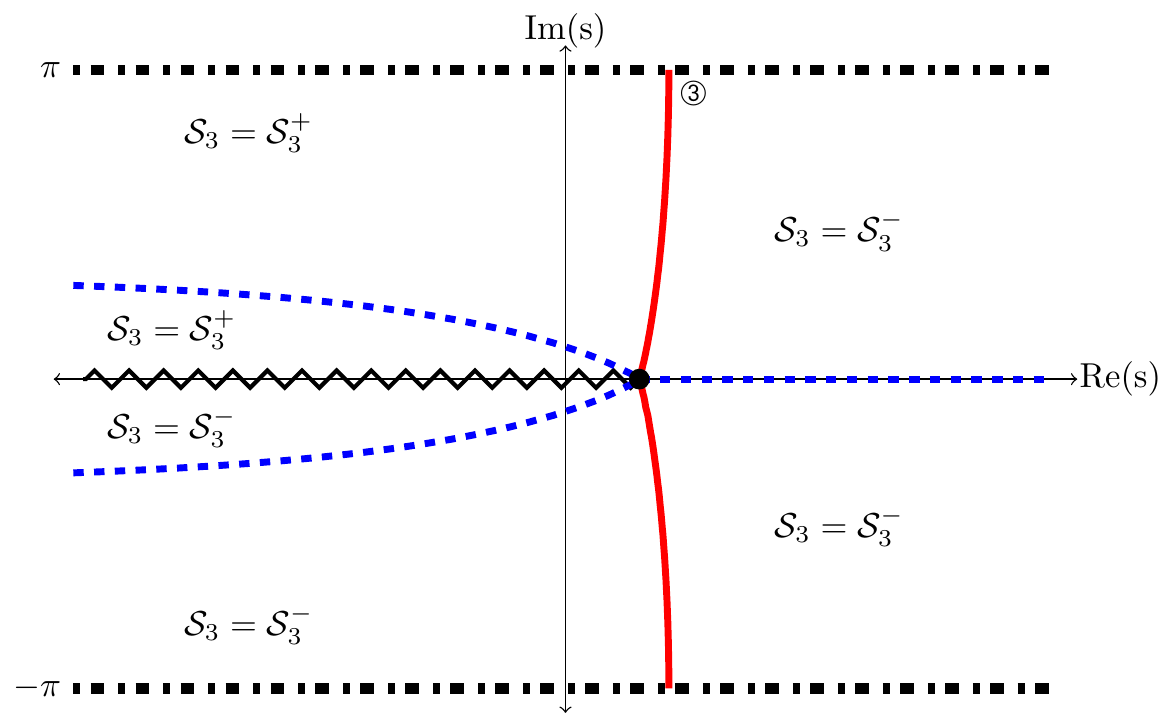}
    \caption{General values of $\mathcal{S}_3$.}\label{Stokes multipliers W03}
  \end{subfigure}
  \hfill
  \begin{subfigure}[b]{0.45\textwidth}\centering
    \includegraphics[scale=0.55]{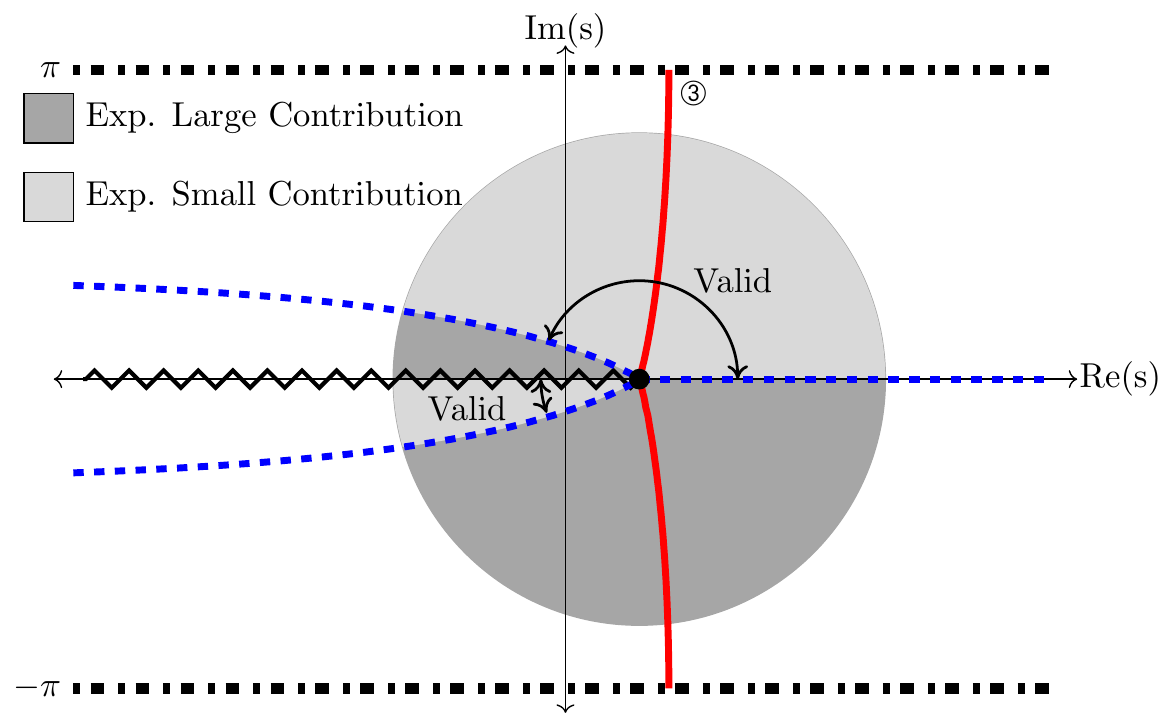}
      \caption{Regions of validity.}\label{Regions of validity W03}
  \end{subfigure}
    \end{adjustbox}
\caption{Figure \ref{Stokes multipliers W03} illustrates the value $\mathcal{S}_3$, which is defined by \eqref{explicit Stokes mult} and \eqref{Stokes multiplier notation defn}, as Stokes curves are crossed. Figure \ref{Regions of validity W03} illustrates the regions of validity of the asymptotic solution described by \eqref{qPI type A complete asym}. The dark and light grey shaded regions denote regions in which the exponential contribution associated with $\chi_3$ is large and small respectively. Hence, the regions of validity are those which are shaded in light grey.}\label{FIRST FIG TOTAL}
\end{figure}

We now determine regions in $\mathcal{D}_0$ in which the asymptotic behaviour of \eqref{qPI rescaled} is described by the power series expansion \eqref{qPI type A complete asym}, referred to as regions of validity. From Figure \ref{qPI singulant behaviours} we observe that the exponential contribution associated with $\chi_3$ is exponentially small in the neighbourhood of the upper Stokes curves since $\text{Re}(\chi_3)>0$, and therefore the presence of $\exp(-\chi_3/\epsilon)$ does not affect the dominance of the leading order behaviour in \eqref{qPI type A complete asym}. Hence, the value of the Stokes multiplier, $\mathcal{S}_3$, in the neighbourhood of the upper Stokes curve may be freely specified, and will therefore contain a free parameter hidden beyond-all-orders. The values of $\mathcal{S}_3$ in the regions of $\mathcal{D}_0$ is illustrated in Figure \ref{Stokes multipliers W03}.

The remainder term associated with $\chi_3$ will exhibit Stokes switching and therefore the value of $\mathcal{S}_3$ varies as it crosses a Stokes curve, say, from state 1 to state 2. Using the naming convention 

\begin{figure}[H]
\centering
\begin{adjustbox}{minipage=\linewidth,scale=1}
  \begin{subfigure}[b]{0.45\textwidth}\centering
  \includegraphics[scale=0.5]{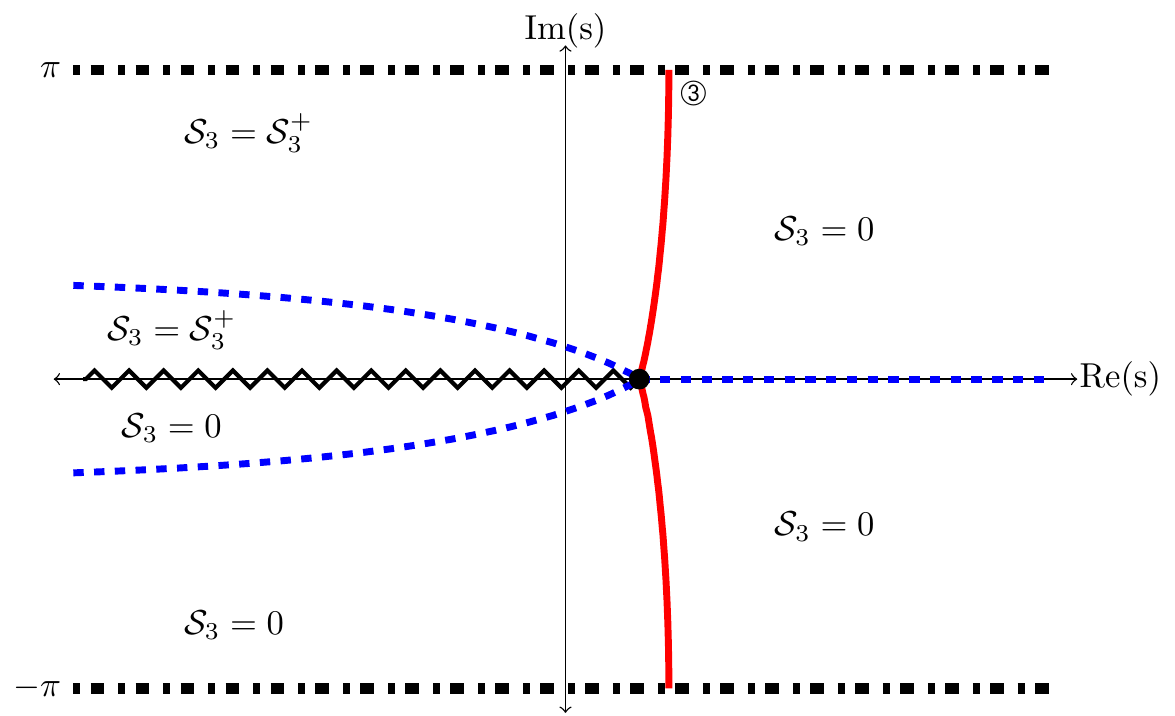}
    \caption{Special values of $\mathcal{S}_3$.}\label{Stokes multipliers special W03}
  \end{subfigure}
  \hfill
  \begin{subfigure}[b]{0.45\textwidth}\centering
    \includegraphics[scale=0.5]{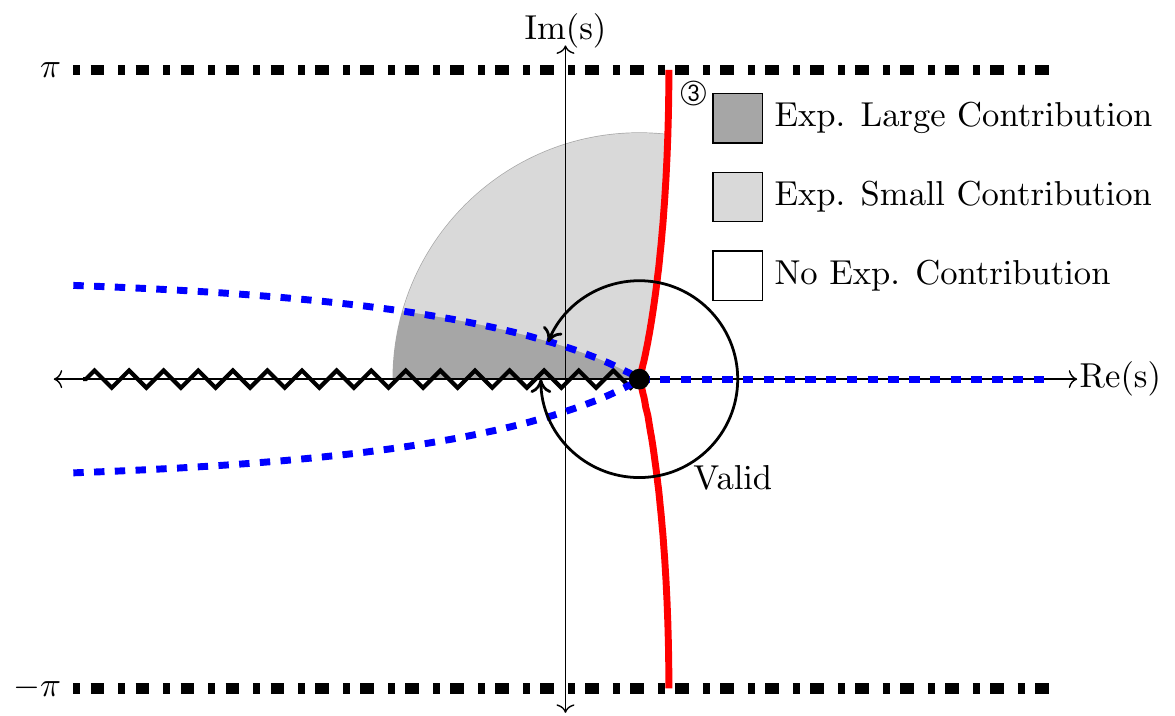}
      \caption{Extended regions of validity.}\label{Regions of validity special W03}
  \end{subfigure}
    \end{adjustbox}
\caption{Figures \ref{Stokes multipliers special W03} and \ref{Regions of validity special W03} correspond to those in Figure \ref{FIRST FIG TOTAL} for the choice of $\mathcal{S}^-_3=0$. This specific choice gives an asymptotic solution of \eqref{qPI rescaled} with an extended region of validity. Non-shaded regions denote regions in which there are no exponential contributions present. The legend in this figure will be used throughout the remainder of this study, unless stated otherwise.}
\end{figure}

\begin{figure}[h!]
\centering
\begin{minipage}[t]{0.45\linewidth} 
\centering
\scalebox{1}{ 
\includegraphics[scale=0.55]{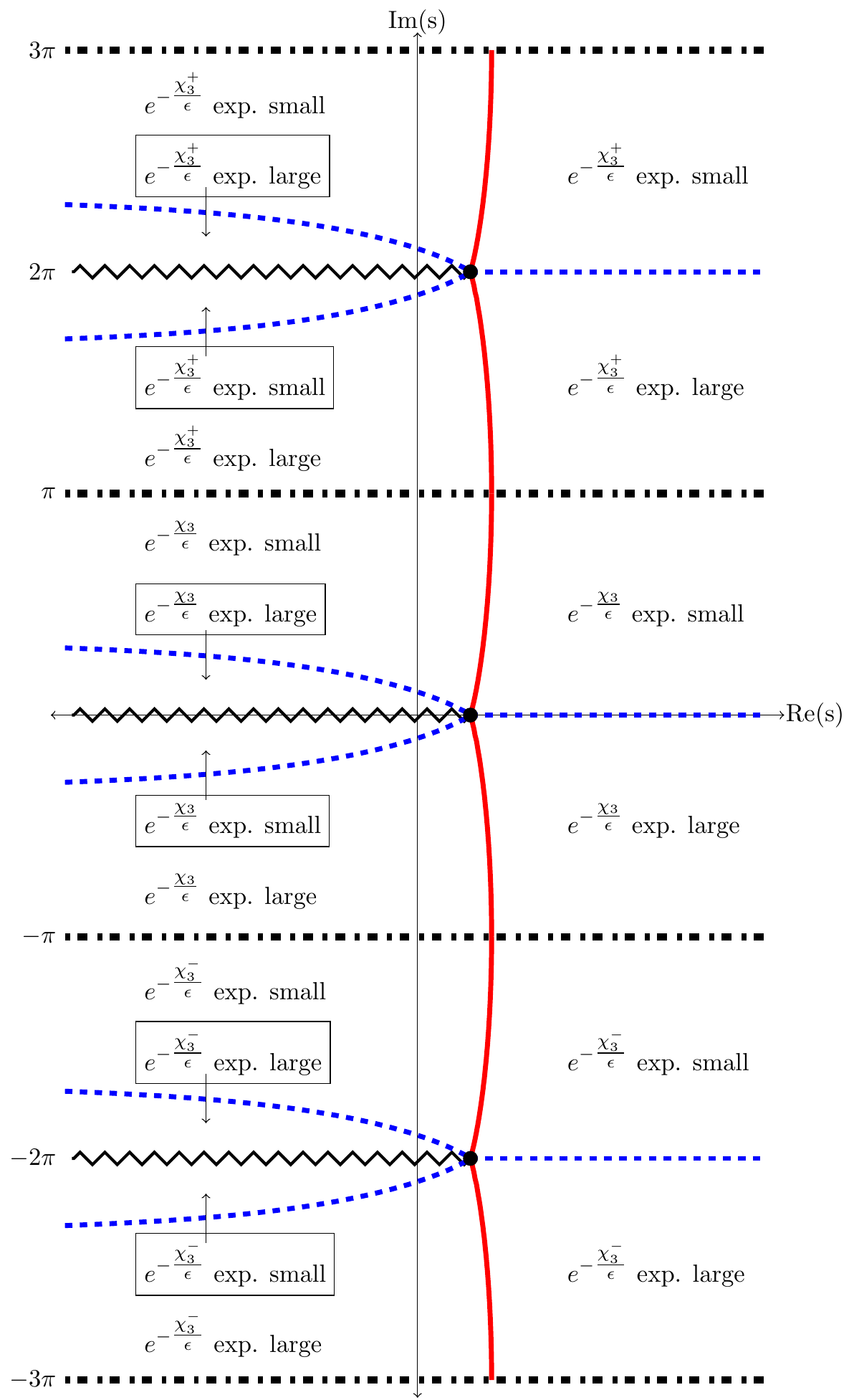}
}
\subcaption{Stokes structure of $\chi_3$ in the domains $\mathcal{D}_0$, $\mathcal{D}_1$ and $\mathcal{D}_{-1}$.}\label{extended stokes struc for chi3}
\end{minipage}
\begin{minipage}[t]{0.45\linewidth} 
\centering
\scalebox{1}{ 
\includegraphics[scale=0.55]{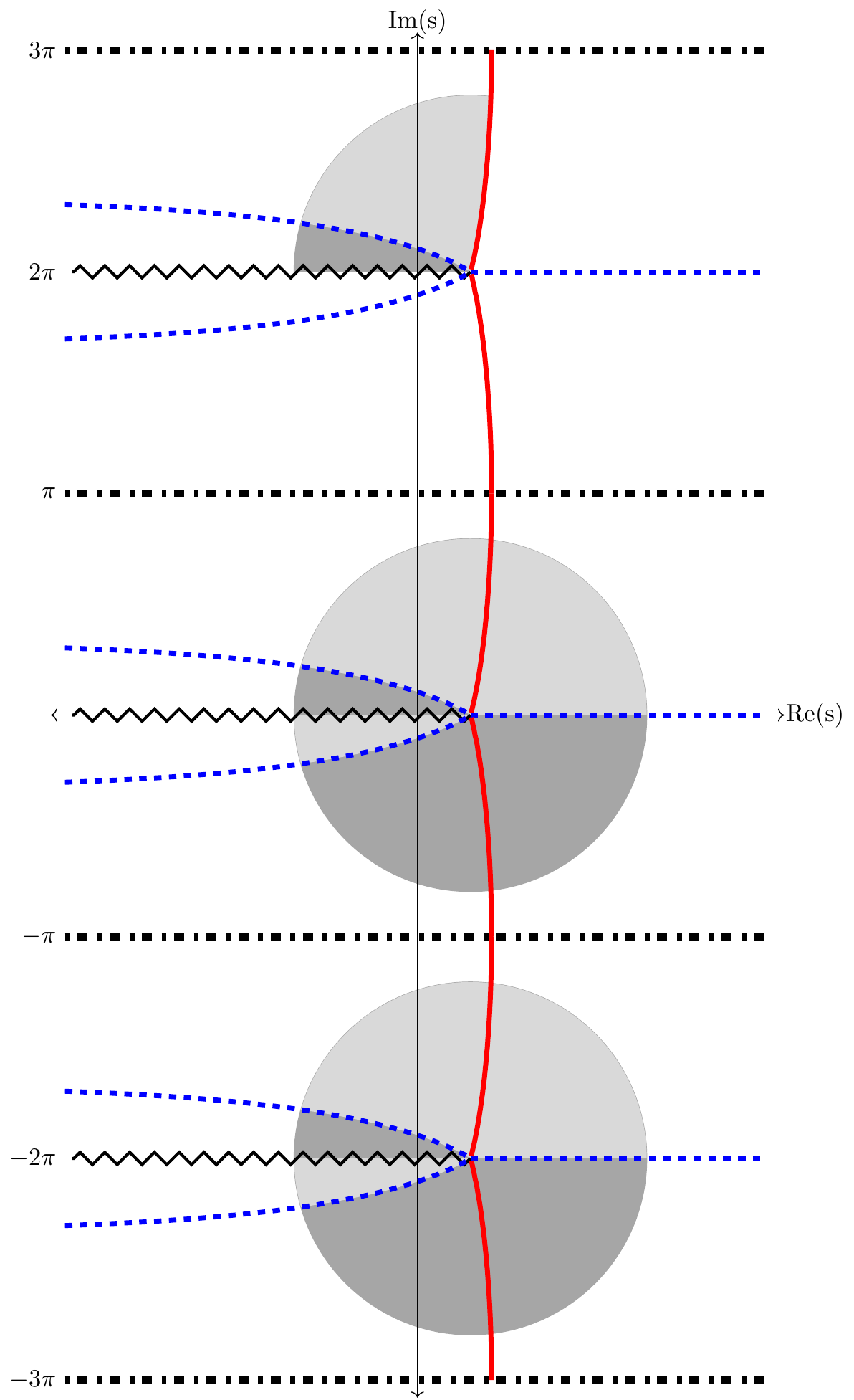}
}
\subcaption{Exponential contributions originating from the domains $\mathcal{D}_0$, $\mathcal{D}_1$ and $\mathcal{D}_{-1}$.}\label{extended stokes struc for chi3 with exponential contributions}
\end{minipage}
\caption{This figure illustrates the Stokes structure depicted in Figure \ref{qPI exponential behaviours} extended to the adjacent domains $\mathcal{D}_1$ and $\mathcal{D}_{-1}$ as described by \eqref{kth adjacent domain}. In fact, the Stokes structure in the $k^{\text{th}}$-adjacent domains are identical to those in $\mathcal{D}_0$ since $W_{0}$ is $2\pi i$-periodic. Each adjacent domain contributes an exponential contribution to the asymptotic solution described by \eqref{qPI type A complete asym}. From figure \ref{extended stokes struc for chi3}, we see that the presence of the exponential contribution from $\mathcal{D}_{-1}$ does not affect the dominance of \eqref{qPI type A complete asym} in $\mathcal{D}_0$ and hence its associated Stokes multiplier may be freely specified. However, the presence of the exponential contribution from $\mathcal{D}_{1}$ dominates the asymptotic solution, \eqref{qPI type A complete asym}. In order for \eqref{qPI type A complete asym} to remain valid in $\mathcal{D}_0$, we require the value of the Stokes multiplier associated with exponential contribution in $\mathcal{D}_1$ must be chosen to be equal to zero. Figure \ref{extended stokes struc for chi3 with exponential contributions} illustrates the regions where each exponential contribution are present.}\label{extended stokes struc for chi3 TOTAL}
\end{figure} 
\noindent described by \eqref{Stokes multiplier notation defn}, we denote these states by $\mathcal{S}^-_3$ and $\mathcal{S}^+_3$ respectively. If we assume that the value of $\mathcal{S}_3$ is nonzero on either side of the upper Stokes curve, then we conclude that the exponentially small contribution associated with $\chi_3$ is present in the regions bounded by the upper anti-Stokes curve, the anti-Stokes curve emanating from the singularity along the positive real $s$-axis and the curve $\text{Im}(s)=\pi$. Furthermore, the exponential contribution associated with $\chi_3$ is also exponentially small in the region bounded by the branch cut and the lower anti-Stokes curve. The regions of validity of the asymptotic solution described by \eqref{qPI type A complete asym} are illustrated in Figure \ref{Regions of validity W03}.

However, for special choices of the free parameter hidden beyond-all-orders, we can obtain asymptotic solutions with an extended range of validity in $\mathcal{D}_0$. If we specify the value of $\mathcal{S}^-_3$ to be equal to zero, then the exponential contribution associated with $\chi_3$ is no longer present in regions where it is normally exponentially large. In this case, the region of validity is extended by two additional adjacent sectorial regions in $\mathcal{D}_0$ as illustrated in Figure \ref{Regions of validity special W03}. We note that the case where $\mathcal{S}^+_3=0$ is specified can also give Type A solutions with an extended region of validity. However, this only extends the regions of validity of \eqref{qPI type A complete asym} by one additional sectorial region. In both cases the value of $\mathcal{S}_3$ is specified, and therefore the asymptotic solution described by \eqref{qPI type A complete asym} is uniquely determined; we call these special Type A asymptotic solutions.

Figure \ref{extended stokes struc for chi3 TOTAL} illustrates the Stokes structure in the adjacent domains $\mathcal{D}_1$ and $\mathcal{D}_{-1}$ as described by \eqref{kth adjacent domain}. Due to the $2\pi i$-periodic nature of $W_{0}$, the Stokes structure is also $2\pi i$-periodic as shown in Figure \ref{extended stokes struc for chi3 TOTAL}. Hence, we obtain an exponential contribution in each adjacent domain $\mathcal{D}_k$. For integers $k\leq-1$, there are exponentially small contributions present in the adjacent domains $\mathcal{D}_k$. The presence of these exponentially small contributions do not affect the asymptotic behaviour in the principal domain, $\mathcal{D}_0$ and hence the corresponding Stokes multipliers may be freely specified. However, for integers $k\geq1$ 
\begin{figure}[h!]
\centering
\begin{adjustbox}{minipage=\linewidth,scale=1}
  \begin{subfigure}[t]{0.45\textwidth}\centering
  \includegraphics[scale=0.5]{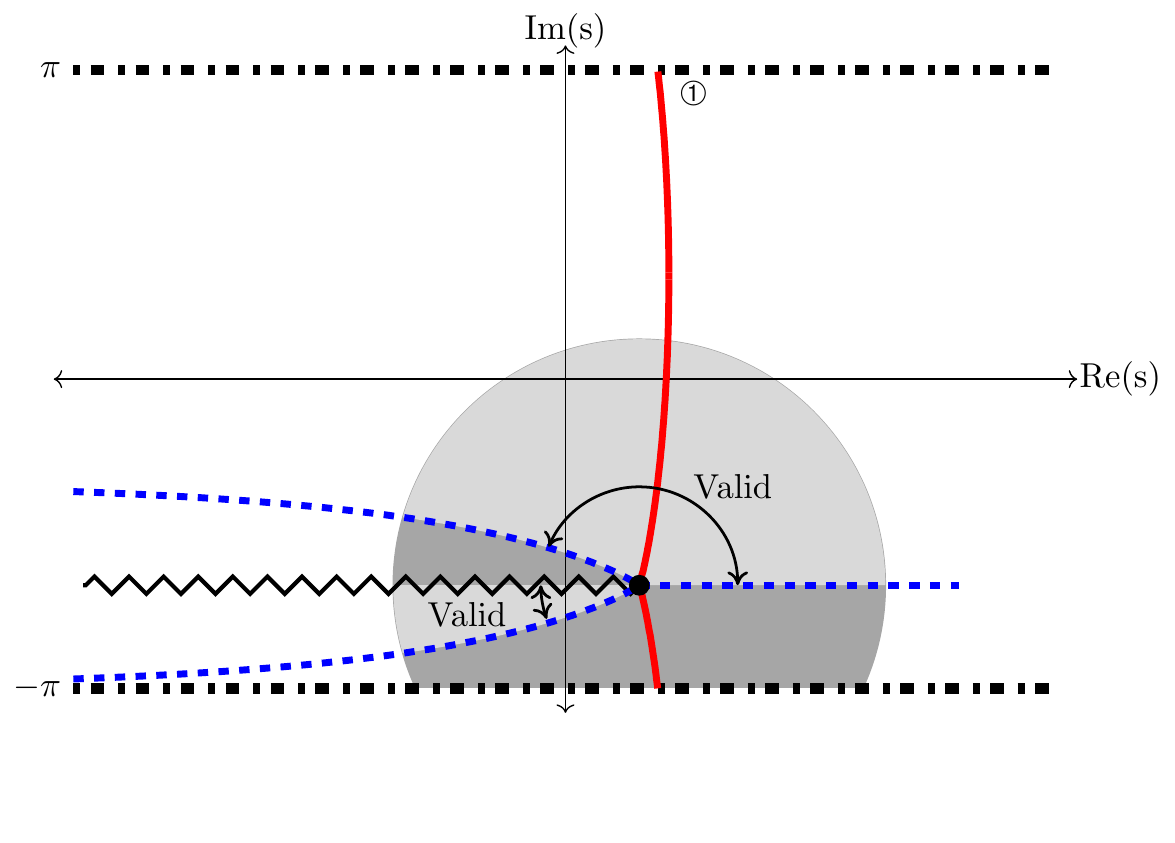}
    \caption{Regions of validity for the asymptotic solution described by \eqref{qPI type A complete asym W01}.}\label{fig type A w01 regions}
  \end{subfigure}
  \hfill
  \begin{subfigure}[t]{0.45\textwidth}\centering
    \includegraphics[scale=0.5]{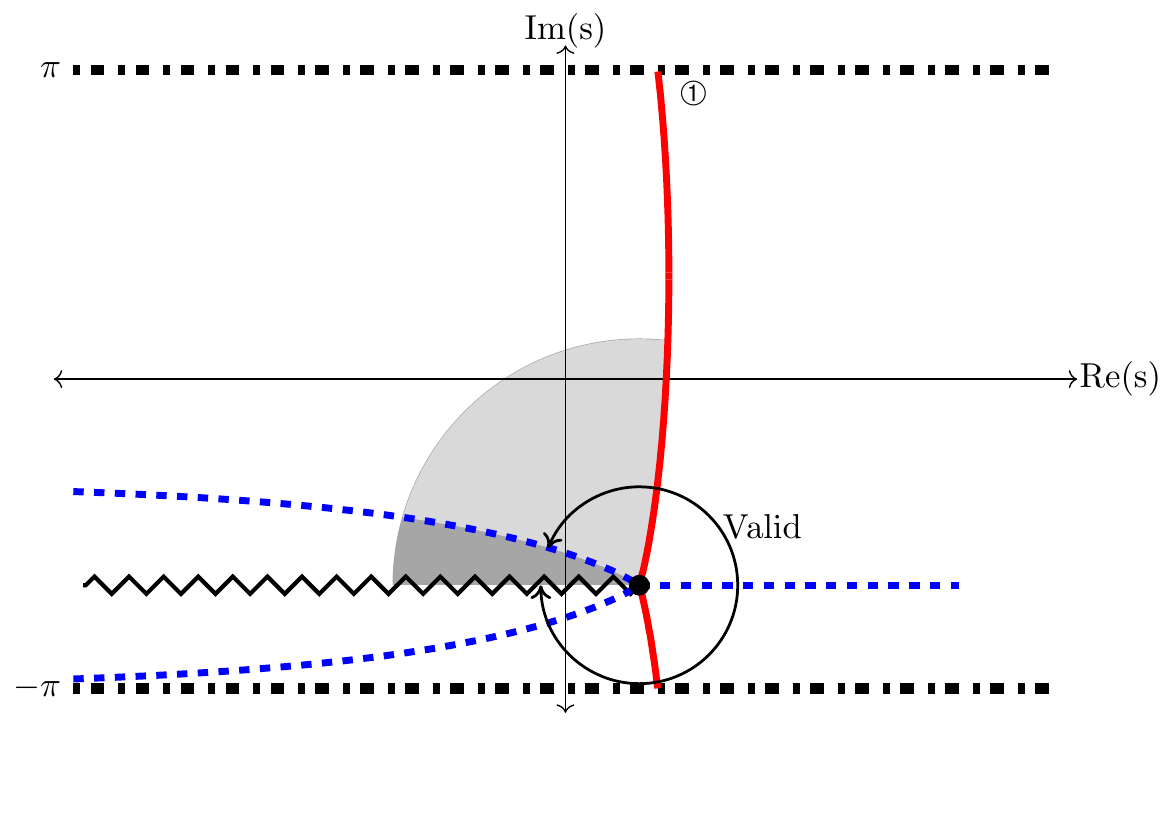}
      \caption{Extended regions of validity for the asymptotic solution described by \eqref{qPI type A complete asym W01}.}\label{fig type A w01 special regions}
  \end{subfigure}  
  \hfill
  \begin{subfigure}[t]{0.45\textwidth}\centering
  \includegraphics[scale=0.5]{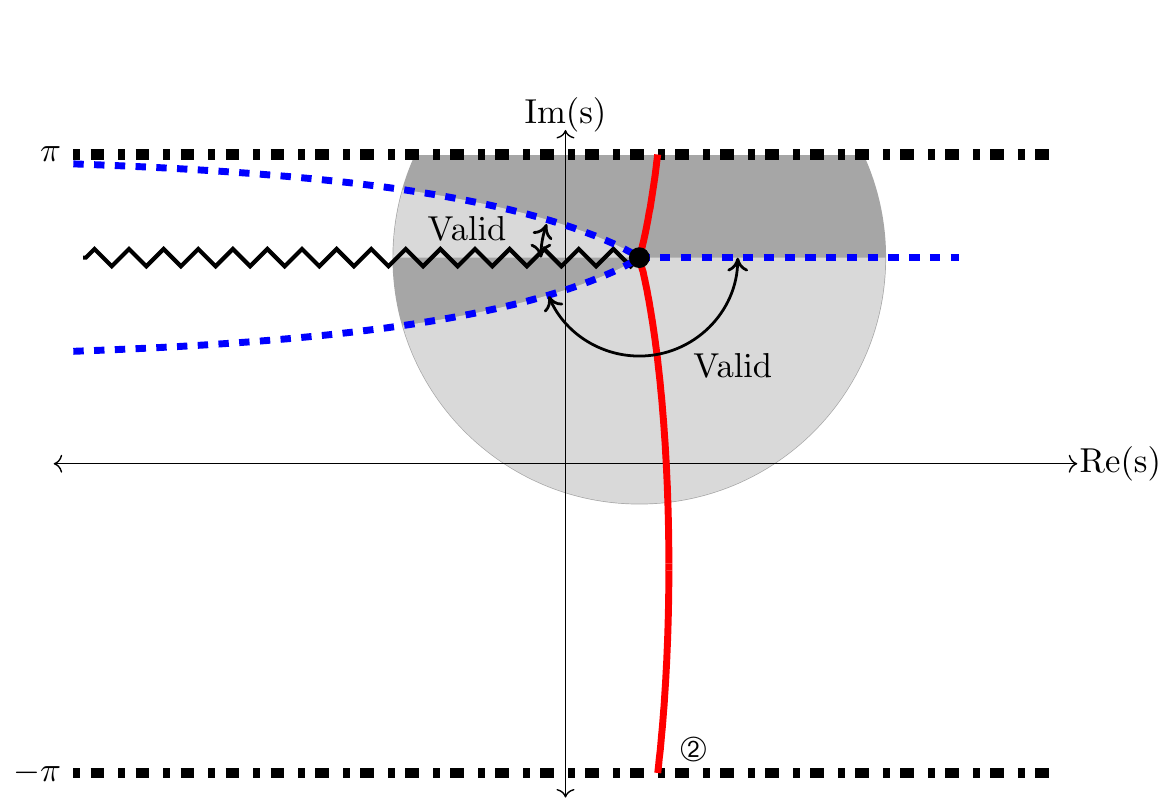}
    \caption{Regions of validity for the asymptotic solution described by \eqref{qPI type A complete asym W02}.}\label{fig type A w02 regions}
  \end{subfigure}
  \hfill
  \begin{subfigure}[t]{0.45\textwidth}\centering
    \includegraphics[scale=0.5]{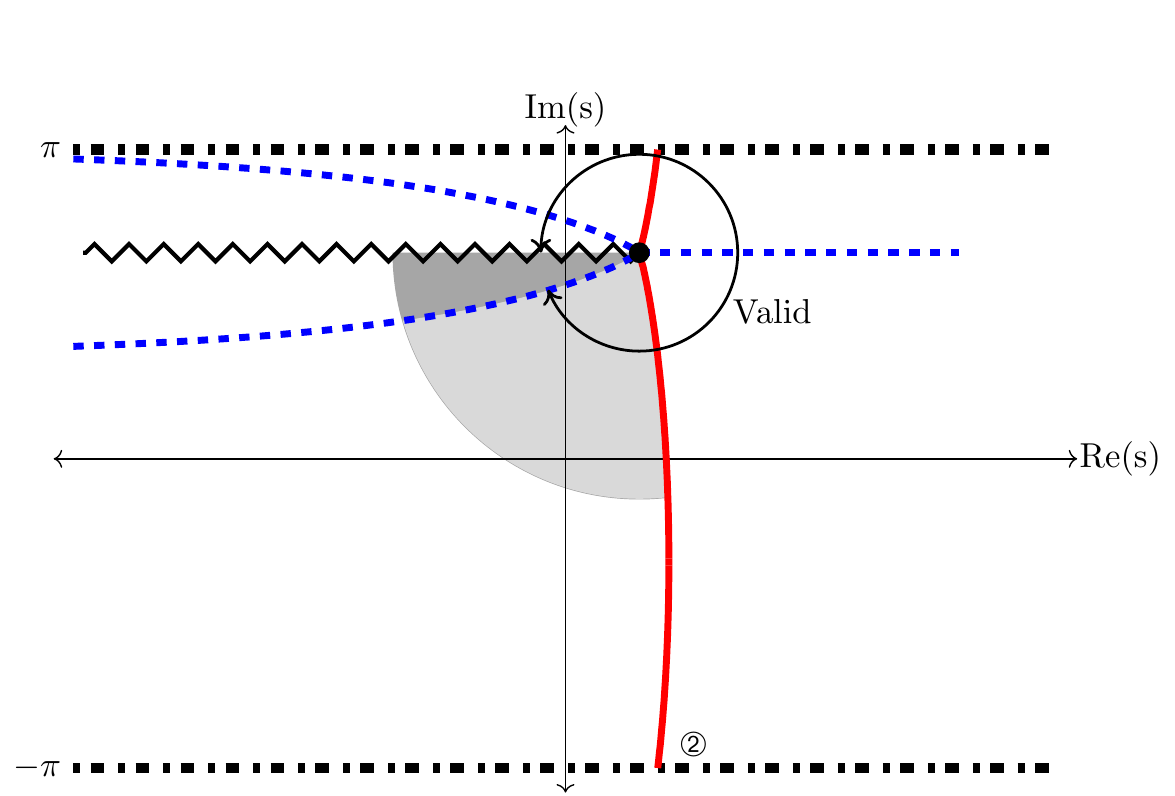}
      \caption{Extended regions of validity for the asymptotic solution described by \eqref{qPI type A complete asym W02}.}\label{fig type A w02 special regions}
  \end{subfigure}
    \end{adjustbox}
\caption{This figure illustrates the Stokes structure and regions of validity for the remaining two Type A solutions. We observe that these Stokes structures are vertical translations of those illustrated in Figure \ref{qPI_stokes TOTAL}, which is due to the symmetry described by \eqref{rescaled qPI symmetry}. Figures \ref{fig type A w01 regions} and \ref{fig type A w01 special regions} illustrate the regions of validity for the general and special asymptotic solution described by \eqref{qPI type A complete asym W01}. Figures \ref{fig type A w02 regions} and \ref{fig type A w02 special regions} illustrate the regions of validity for the general and special asymptotic solution described by \eqref{qPI type A complete asym W02}.}\label{fig remaining type A stokes structure}
\end{figure}
the exponential contributions originating from the adjacent domains $\mathcal{D}_k$ do affect the asymptotic behaviour in $\mathcal{D}_0$. In order for the asymptotic solution \eqref{qPI type A complete asym} to correctly describe the solution behaviour in $\mathcal{D}_0$, the value of the Stokes multipliers must be specified such that the are not present in $\mathcal{D}_0$. The presence of the exponential contributions in the domains $\mathcal{D}_{-1}, \mathcal{D}_{0}$ and $\mathcal{D}_{1}$ is illustrated in Figure \ref{extended stokes struc for chi3 with exponential contributions}.

\begin{figure}[H]
\centering
\begin{adjustbox}{minipage=\linewidth,scale=0.9}
  \begin{subfigure}[b]{0.3\textwidth}
  \includegraphics[scale=0.47]{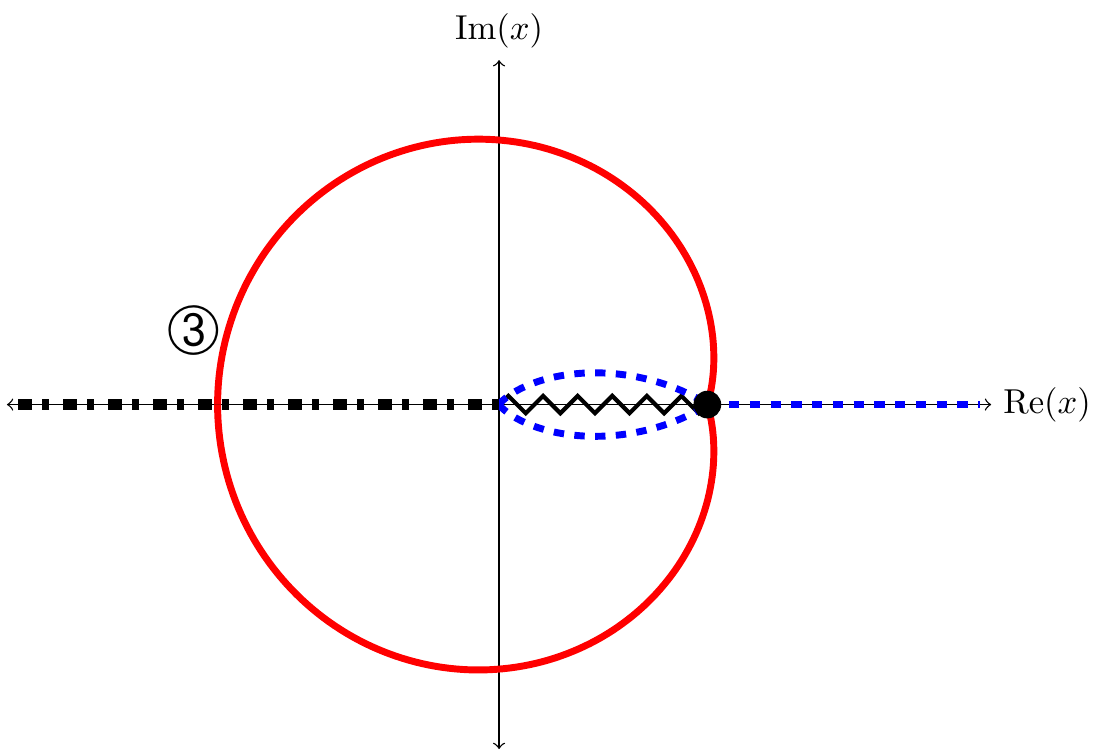}
    \caption{Stokes structure of \eqref{qPI type A complete asym} in the $x$-plane.}\label{stokes struc w03 x plane}
  \end{subfigure}
  \hfill
  \begin{subfigure}[b]{0.3\textwidth}
    \includegraphics[scale=0.47]{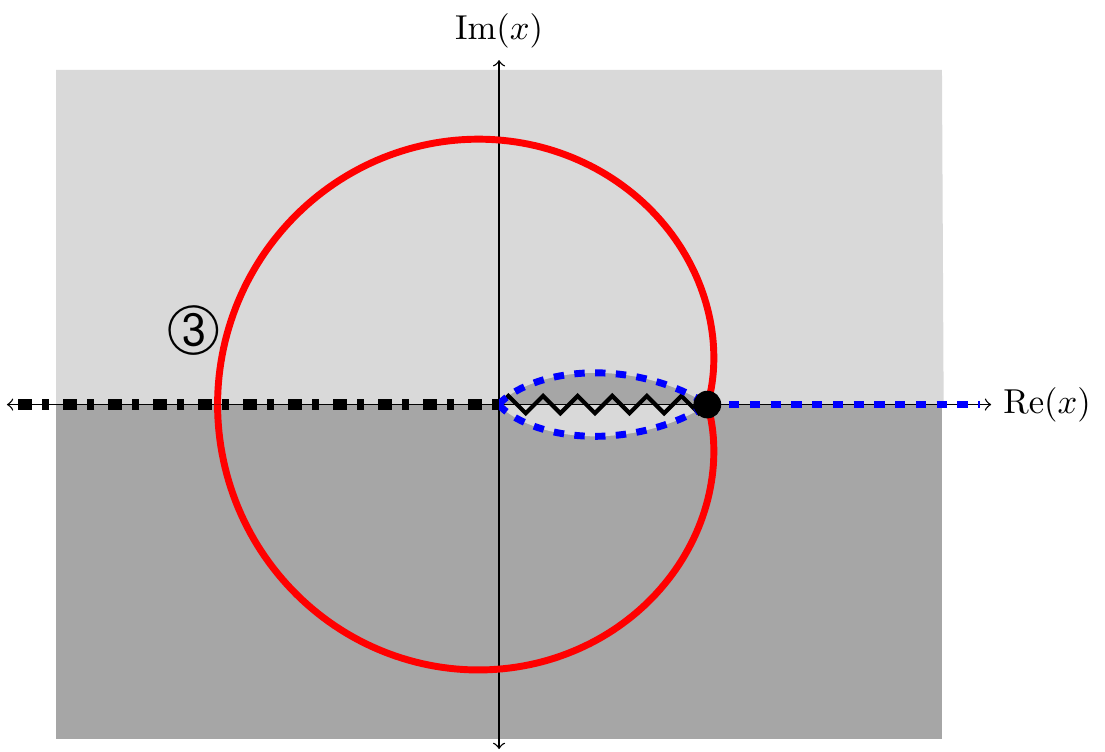}
      \caption{Regions of validity of \eqref{qPI type A complete asym} in the $x$-plane.}\label{stokes struc w03 x plane regions of valid}
  \end{subfigure}
  \hfill
  \begin{subfigure}[b]{0.3\textwidth}
  \includegraphics[scale=0.47]{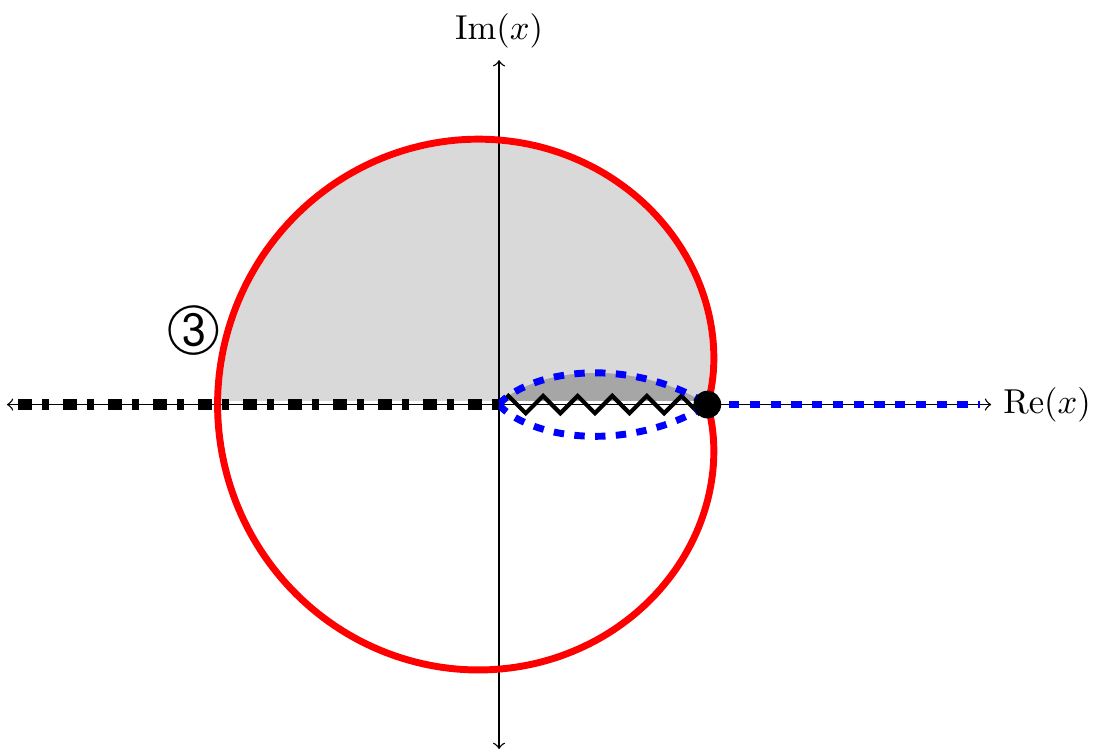}
    \caption{Extended regions of validity of \eqref{qPI type A complete asym} in the $x$-plane.}\label{stokes struc w03 x plane regions of valid special}
  \end{subfigure}
  \hfill 
  \begin{subfigure}[b]{0.3\textwidth}
  \includegraphics[scale=0.47]{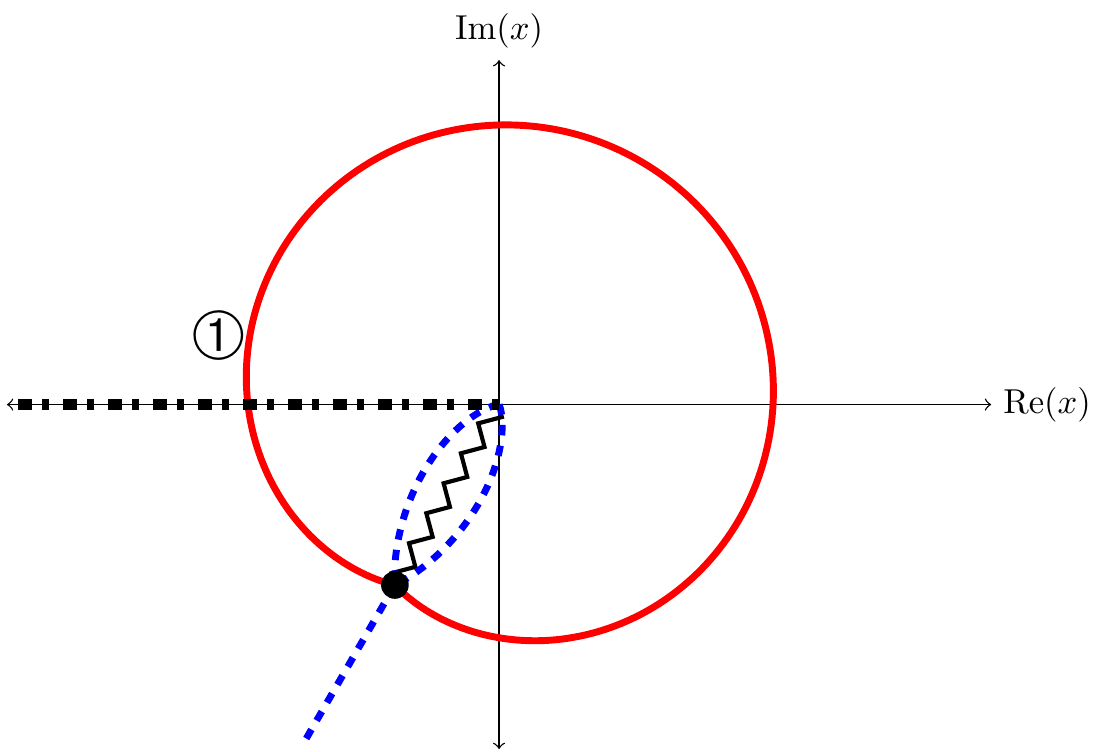}
    \caption{Stokes structure of \eqref{qPI type A complete asym W01} in the $x$-plane.}\label{stokes struc w01 x plane}
  \end{subfigure}
  \hfill
  \begin{subfigure}[b]{0.3\textwidth}
    \includegraphics[scale=0.47]{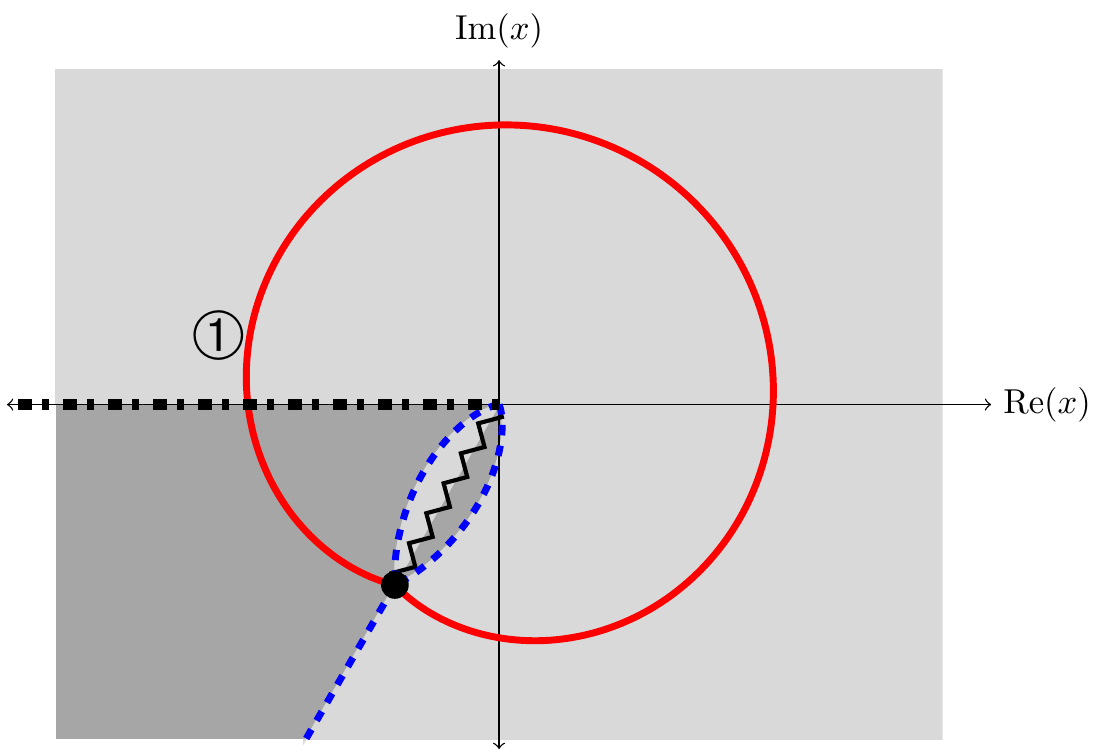}
      \caption{Regions of validity of \eqref{qPI type A complete asym W01} in the $x$-plane.}\label{stokes struc w01 x plane regions of valid}
  \end{subfigure}
  \hfill
  \begin{subfigure}[b]{0.3\textwidth}
  \includegraphics[scale=0.47]{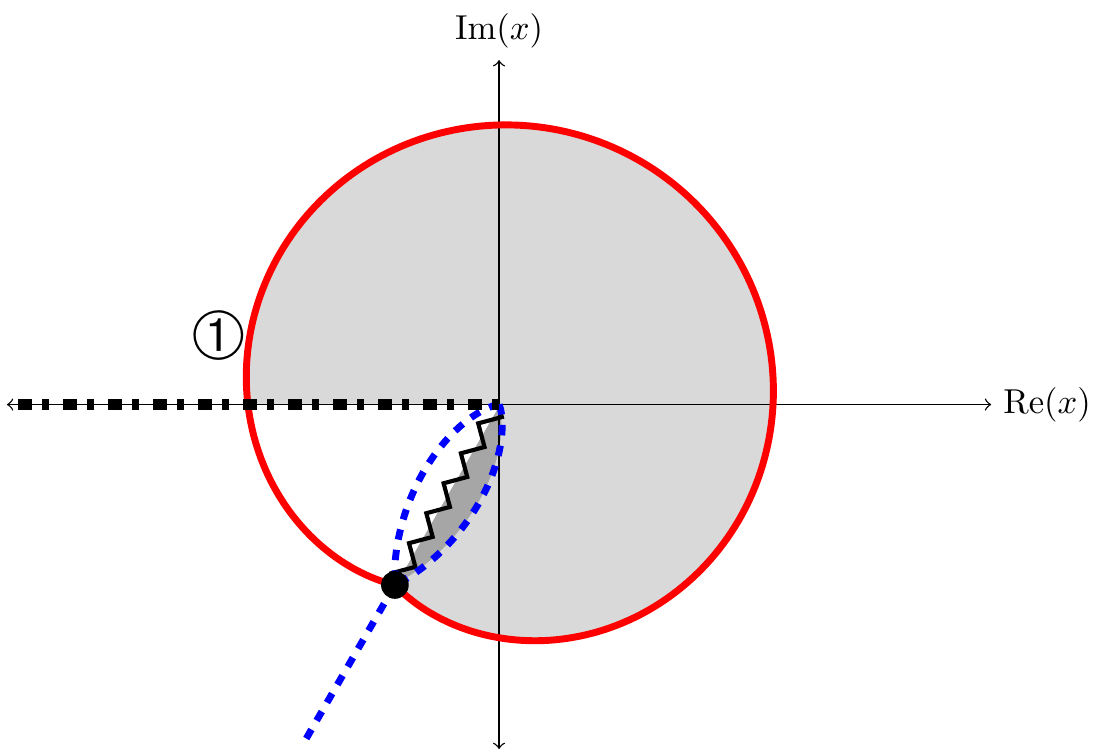}
    \caption{Extended regions of validity of \eqref{qPI type A complete asym W01} in the $x$-plane.}\label{stokes struc w01 x plane regions of valid special}
  \end{subfigure} 
  \hfill 
  \begin{subfigure}[b]{0.3\textwidth}
  \includegraphics[scale=0.47]{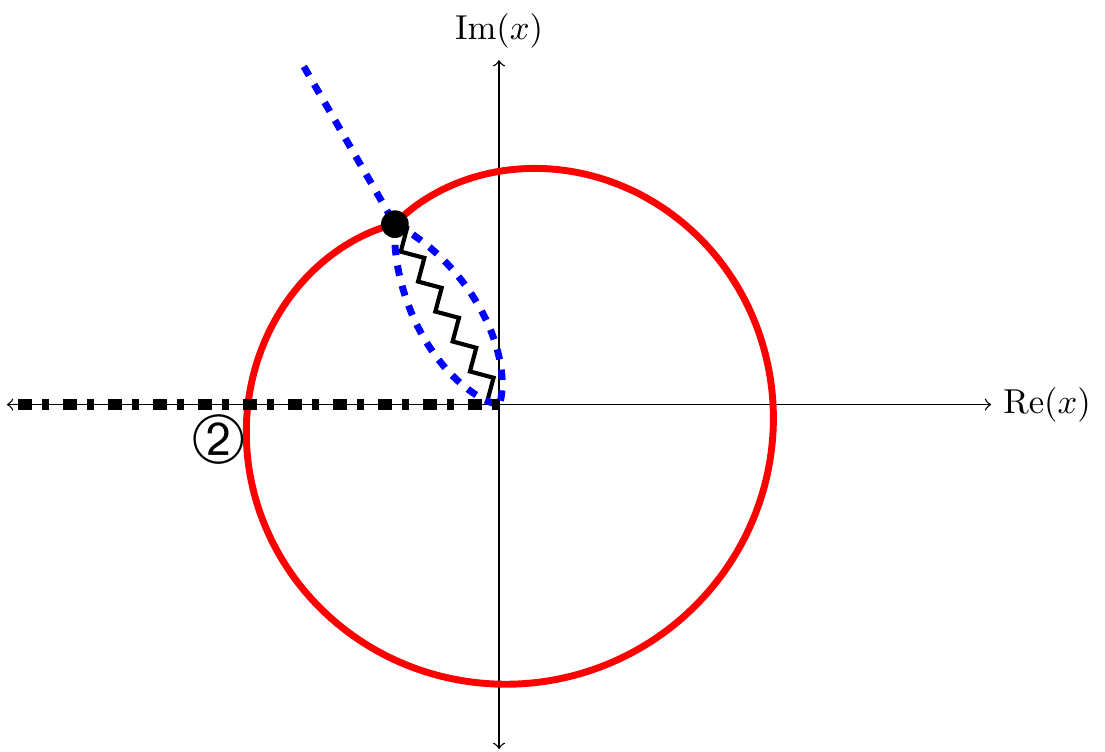}
    \caption{Stokes structure of \eqref{qPI type A complete asym W02} in the $x$-plane.}\label{stokes struc w02 x plane}
  \end{subfigure}
  \hfill
  \begin{subfigure}[b]{0.3\textwidth}
    \includegraphics[scale=0.47]{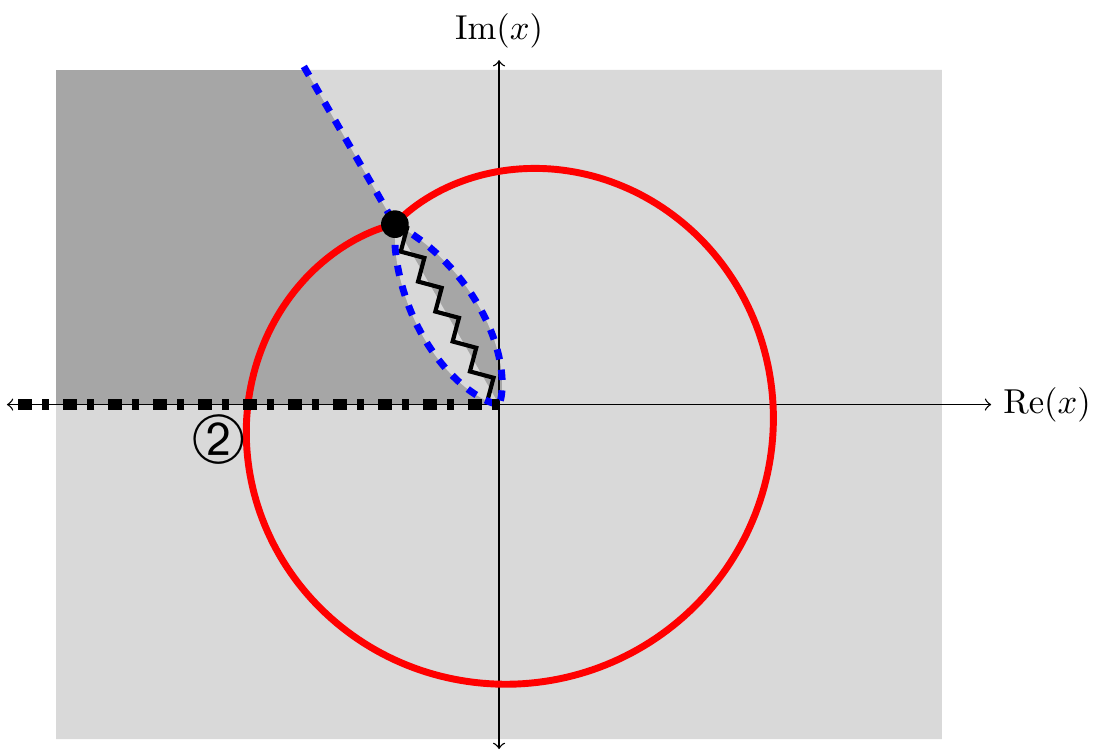}
      \caption{Regions of validity of \eqref{qPI type A complete asym W02} in the $x$-plane.}\label{stokes struc w02 x plane regions of valid}
  \end{subfigure}
  \hfill
  \begin{subfigure}[b]{0.3\textwidth}
  \includegraphics[scale=0.47]{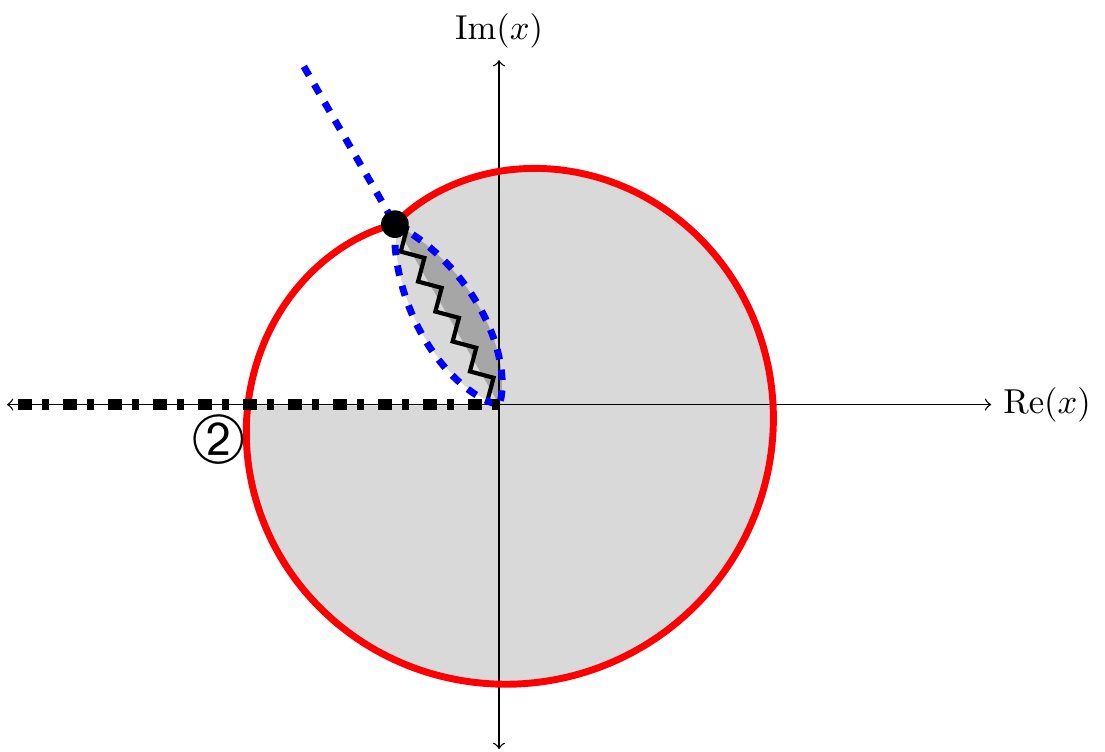}
    \caption{Extended regions of validity of \eqref{qPI type A complete asym W02} in the $x$-plane.}\label{stokes struc w02 x plane regions of valid special}
  \end{subfigure}   
    \end{adjustbox}
\caption{This figure illustrates the Stokes structure of Type A solutions \eqref{qPI type A complete asym}-\eqref{qPI type A complete asym W02} of \eqref{qPI rescaled} in the complex $x$-plane with $q=1+0.2i$. Under the inverse transformation \eqref{qPI reverse transformation}, the Stokes and anti-Stokes curves are now described by $q$-spirals in the $x$-plane. The branch cuts of the singulants, $\chi$, in the $x$-plane extend from the singularities and terminate at the origin. As we are using the leading order term of the inverse transformation in the limit $\epsilon\rightarrow 0$, the boundaries of $\mathcal{D}_0$ are mapped to the negative real $x$ axis, which corresponds to a logarithmic branch cut. Figure \ref{stokes struc w03 x plane regions of valid} illustrates the region of validity (light grey shaded regions) of \eqref{qPI type A complete asym} which contain a free parameter hidden beyond-all-orders. The regions of validity of \eqref{qPI type A complete asym} for which $\mathcal{S}^-_3$ is uniquely specified are illustrated in Figure \ref{stokes struc w03 x plane regions of valid special}. The corresponding Stokes structure and regions of validity for the remaining two Type A solutions are illustrated in Figures \ref{stokes struc w01 x plane}-\ref{stokes struc w02 x plane regions of valid special}.}\label{stokes struc w03 x plane total}
\end{figure}

The corresponding analysis of the Stokes structure and switching behaviour of the exponential contributions associated with the asymptotic solutions \eqref{qPI type A complete asym W01} and \eqref{qPI type A complete asym W02} may be obtained by using the symmetry \eqref{rescaled qPI symmetry}. Consequently, the Stokes structure associated with $\chi_1$ and $\chi_2$ are vertical translations of the $\chi_3$-Stokes structure by $\mp 2i \pi$, respectively. The corresponding figures are contained in Figure \ref{fig remaining type A stokes structure}.

As the Stokes structure and switching behaviour of the exponential contributions have been determined in the domain $\mathcal{D}_0$, we may finally determine the Stokes structure in the original complex $x$-plane. In order to determine the Stokes structure in the $x$-plane we reverse the scaling transformations. In particular, from the scalings given by \eqref{q difference rescale independent variable} we find that the leading order mapping from $s$ to $x$ is given by
\begin{equation}\label{qPI reverse transformation}
s=\frac{\epsilon\log(x)}{\log(1+\epsilon)}\sim \log(x) +\mathcal{O}(\epsilon),
\end{equation}
as $\epsilon\rightarrow 0$. 

We illustrate the Stokes structures of Type A solutions for the choice of $q=1+0.2i$. Using \eqref{qPI reverse transformation}, the singulants, $\chi_j(s)$, can be written as a function of $x$. We then compute the Stokes structure in the complex $x$-plane using \textsc{Matlab}. The corresponding Stokes structure for the asymptotic solution described by \eqref{qPI type A complete asym} in the complex $x$-plane is illustrated in Figure \ref{stokes struc w03 x plane}. In these figures, the (anti-) Stokes curves and branch cuts in the complex $x$-plane follow the convention illustrated in Figure \ref{qPI_stokes TOTAL}.

In particular, the complex $x$-plane contains a logarithmic branch cut (dot-dashed curve) along the negative real $x$ axis as a result of using the leading order term of the inverse transformation described by \eqref{qPI reverse transformation}. In fact, the boundaries of $\mathcal{D}_0$ are mapped to this branch cut in the complex $x$-plane. The inverse transformation maps the Stokes and anti-Stokes curves in the $s$-plane to $q$-spirals in the complex $x$-plane. From Figure \ref{stokes struc w03 x plane total}, we find the Stokes and anti-Stokes curve separate the complex $x$-plane into sectorial regions bounded by arcs of spirals.

\section{Type B Asymptotics}\label{S:Type B Asymptotics}
In this section we investigate Type B solutions of \eqref{qPI rescaled}. These are the asymptotic solutions of \eqref{qPI rescaled}, which are described by $W_{0,4}$ to leading order as $\epsilon\rightarrow 0$. The analysis involved in the subsequent sections is nearly identical to Sections \ref{S:Asymptotic series expansions} and \ref{S:Exponential Asymptotics}. Hence, we will omit the details and only provide the key results.

Type B solutions may be expanded as a power series in $\epsilon$ of the form
\begin{equation}\label{type b asym series}
W(s)\sim W_{0,4}(s)+\sum_{r=1}^{\infty}\epsilon^ry_r(s),
\end{equation}
as $\epsilon\rightarrow 0$. Following the analysis for Type A solutions, it can be shown that the behaviour of $y_r$ is also described by a factorial-over-power form. The behaviour of $y_r$ is therefore described by
\begin{equation*}\label{type B late order terms}
y_{r}(s)\sim \frac{Y(s)\Gamma(r+\nu)}{\eta(s)^{r+\nu}},
\end{equation*}
as $r\rightarrow\infty$. Recall that the main difference between Type A and Type B solutions is that Type B solutions are singular at three distinct points in $\mathcal{D}_0$ rather than one. This feature will therefore be encoded in the calculation of the singulant function, $\eta$. 

Since the leading order behaviour $W_{0,4}$ is singular at $s_{0,1}, s_{0,2}$ and $s_{0,3}$, we obtain three singulant contributions, which we denote by $\eta_j(s)$. This feature can be deduced from equation \eqref{qPI w04 expression in terms of others}, which shows that $W_{0,4}$ is expressible as the sum of $W_{0,1}, W_{0,2}$ and $W_{0,3}$. Following the analysis in Section \ref{S:Late-order terms}, $\eta_j(s)$ is given by
\begin{equation}\label{qPI type B singulant}
\eta_j(s) = \int_{s_{0,j}}^{s}\cosh^{-1}(\sigma(t))dt,
\end{equation}
for $j=1,2,3$ and where $\sigma$ is given in \eqref{qPI singulant expressions} with $W_{0,3}$ replaced by $W_{0,4}$. Hence we obtain three contributions for $\eta$. Using the results for the late-order terms found in Section \ref{S:Late-order terms}, we find that the late-order terms of \eqref{type b asym series} is given by
\begin{equation}\label{qPI type B late order terms}
y_{r}(s)\sim \sum_{j=1}^{3}\frac{Y_j(s)\Gamma(r+\nu)}{\eta_j(s)^{r+\nu}},
\end{equation}
as $r\rightarrow\infty$, and where $Y_j(s)$ are the prefactor terms associated with $\eta_j(s)$. Hence the asymptotic expansion of Type B solutions of \eqref{qPI rescaled} is given by
\begin{equation}\label{qPI type B asym series with late order terms}
W(s)\sim W_{0,4}(s)+\sum_{j=1}^{3}\sum_{r=1}^{\infty}\frac{\epsilon^{r}Y_j(s)\Gamma(r+\nu)}{\eta_j(s)^{r+\nu}},
\end{equation}
as $\epsilon\rightarrow 0$. As there are three distinct singulant terms in \eqref{qPI type B asym series with late order terms} there will be three subdominant exponentials present (after optimal truncation) and hence Type B solutions will also display Stokes behaviour. 

By applying the Stokes smoothing technique demonstrated in Section \ref{S:Stokes smoothing} to \eqref{qPI type B asym series with late order terms}, the expression which captures the Stokes behaviour of the subdominant exponential correction terms is given by
\begin{equation}\label{type b asym series combination complete}
W(s)\sim W_{0,4}+\sum_{j=1}^{3}\sum_{r=1}^{2N_{\text{opt}}-1}\frac{\epsilon^{r}Y_j(s)\Gamma(r+\nu)}{\eta_j(s)^{r+\nu}}+\sum_{j=1}^{3}\mathcal{\hat{S}}_j(s)\hat{\phi}(s)Y_j(s)e^{-\eta_j(s/\epsilon)},
\end{equation}
as $\epsilon\rightarrow 0$, and where $N_\text{opt}$ is the optimal truncation point. In particular, the Type B prefactor terms satisfy equation \eqref{qPI EXACT PREFACTOR} with $W_0$ replaced by $W_{0,4}$. Similarly, the Stokes multipliers $\mathcal{\hat{S}}(s)$ satisfy \eqref{qPI Stokes multiplier integral} with $W_{0}$ and $\chi$ replaced by $W_{0,4}$ and $\eta$ respectively. In view of the formula \eqref{qPI w04 expression in terms of others}, the leading order behaviour of the Type B solution is a composition of the leading order behaviours of Type A solutions. Consequently, the Stokes behaviour present in this solution will be more complicated as the Stokes curves emanate from more than one point as this allows the possibility of interaction effects. In order to determine the Stokes structure of Type B solutions, we analyze the singulant \eqref{qPI type B singulant}.

\subsection{Stokes Structure}\label{S:Stokes structure type B}
\begin{figure}[h!]
\centering
\begin{adjustbox}{minipage=\linewidth,scale=0.95}
  \begin{subfigure}[t]{0.45\textwidth}\centering
  \includegraphics[scale=0.7]{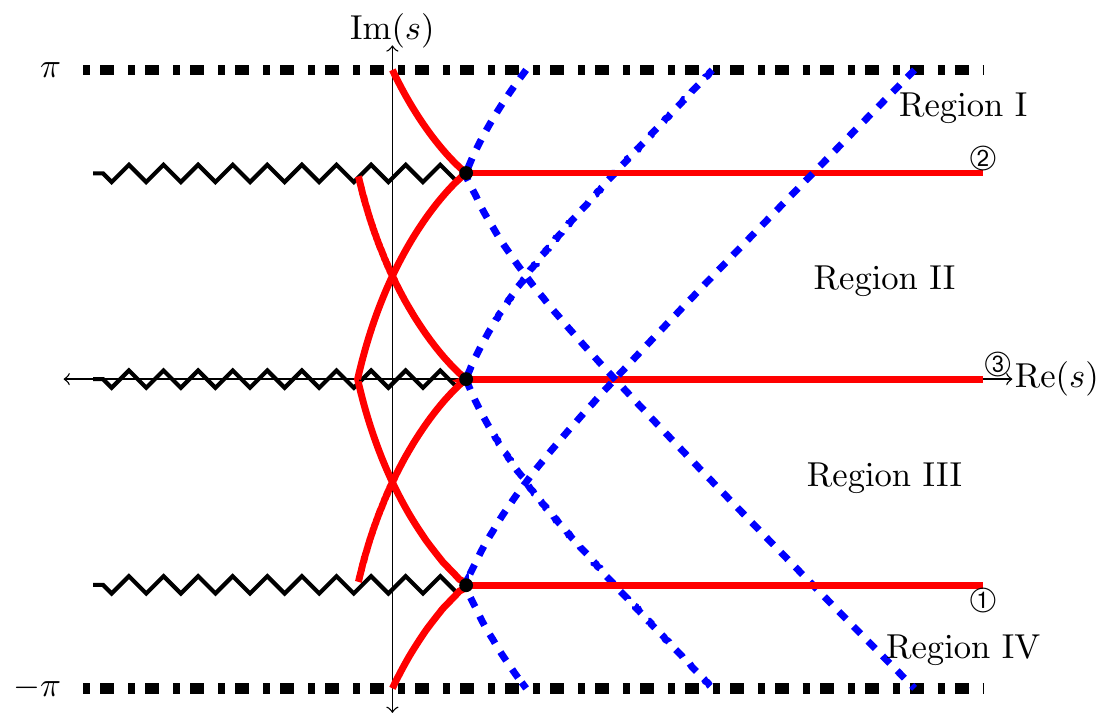}
    \caption{Stokes structure of \eqref{type b asym series combination complete} in the complex $s$-plane.}\label{typeB stokes struc s place zoomout}
  \end{subfigure}
  \hfill
  \begin{subfigure}[t]{0.45\textwidth}\centering
    \includegraphics[scale=0.7]{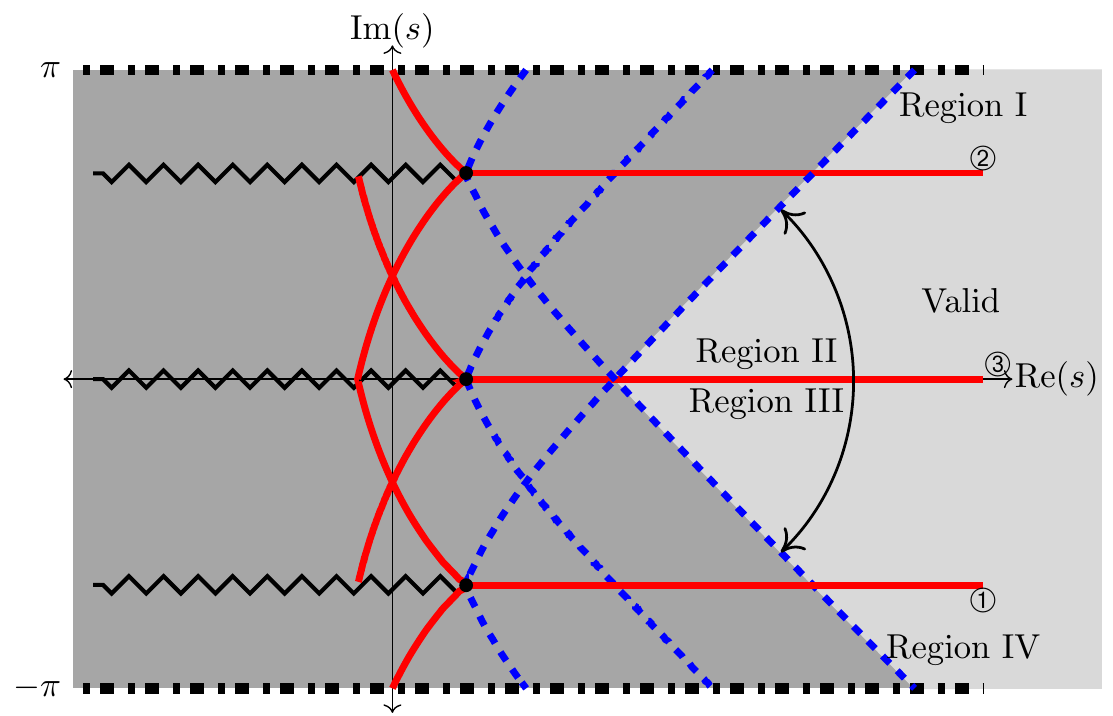}
      \caption{Behaviour of $e^{-\eta_j/\epsilon}$ in the complex $s$-plane.}\label{typeB regions of valid s plane}
  \end{subfigure}  
    \end{adjustbox}
\caption{Stokes structure of the series solution \eqref{type b asym series combination complete} of \eqref{qPI rescaled} in the domain $\mathcal{D}_0$. Regions I-IV denote regions in which the exponential contributions associated with $\eta_j$, which is given by \eqref{qPI type B singulant}, are exponentially small and therefore denote the regions of validity for \eqref{type b asym series combination complete}.}\label{stokes struc w04 s plane}
\end{figure}

The Stokes structure of Type B asymptotic solutions in $\mathcal{D}_0$ is illustrated in Figure \ref{stokes struc w04 s plane}. In Figure \ref{stokes struc w04 s plane} we see that there are three Stokes and two anti-Stokes curves curves emanating from each of the singularities, $s_{0,j}$, for $j=1,2,3$. The Stokes structure for Type B solutions is more complicated as there are Stokes curves which cross into the branch cuts (zig-zag curves) of $\eta_j$ as illustrated in Figure \ref{stokes struc w04 s plane}. As these Stokes curves continue onto another Riemann sheet of $\eta_j$, they may be subject to possible interaction effects from singularities originating in these Riemann sheets. 

However, we observe from Figure \ref{stokes struc w04 s plane} that there are regions which do not contain the Stokes curves continuing into the different Riemann sheets of $\eta_j$. These regions are labelled as regions I-IV in Figure \ref{stokes struc w04 s plane}. We first note that the Stokes curves emanating from the singularities $s_{0,1}, s_{0,2}$ and $s_{0,3}$ asymptote to infinity as $\text{Re}(s)\rightarrow\infty$ as illustrated in Figure \ref{typeB stokes struc s place zoomout}. In Figure \ref{typeB stokes struc s place zoomout}, we find that region I is the region bounded by the upper boundary of $\mathcal{D}_0$, the upper anti-Stokes curve emanating from $s_{0,1}$ and the Stokes curve emanating from $s_{0,2}$, which is labelled by \ding{193}; region II is the region bounded by the upper anti-Stokes curve emanating from $s_{0,1}$, and the Stokes curves labelled by \ding{193} and \ding{194}; region III is the region bounded by the lower anti-Stokes curve emanating from $s_{0,2}$, and the Stokes curves labelled by \ding{192} and \ding{194}; and finally region IV is the region bounded by the lower boundary of $\mathcal{D}_0$, the lower anti-Stokes curve emanating from $s_{0,2}$ and the Stokes curve emanating from $s_{0,1}$, which is labelled by \ding{192}. 

Furthermore, $\text{Re}(\eta_j)$ is positive in each of these four regions and hence the exponential contributions associated with $\eta_j$ are exponentially small there. This is illustrated by the light grey shaded regions in Figure \ref{typeB regions of valid s plane}. We therefore restrict our analysis to regions I-IV as these are the regions in which the dominant asymptotic behaviour is described by \eqref{type b asym series combination complete}. Consequently, the regions of validity of Type B solutions are the regions bounded by the upper and lower anti-Stokes curves emanating from the singularities, $s_{0,1}$ and $s_{0,2}$, respectively, and the boundaries of $\mathcal{D}_0$ containing the positive real $s$ axis. This is the union of regions I-IV and is illustrated in Figure \ref{typeB regions of valid s plane}.

In order to determine Stokes behaviour present in the asymptotic solution \eqref{type b asym series combination complete} we investigate the behaviour of $\eta_j$ in regions I-IV. In Figure \ref{stokes struc w04 s plane}, regions I-IV denotes those of $\mathcal{D}_0$ in which $\text{Re}(\eta_j)>0$. Additionally, the imaginary parts of $\eta_j$ for $j=1,2,3$ are all positive in region I, whereas they are all negative in region IV. Furthermore, we have $\text{Im}(\eta_1)>0$, $\text{Im}(\eta_2)<0$, $\text{Im}(\eta_3)>0$ in region II while $\text{Im}(\eta_1)>0$, $\text{Im}(\eta_2)<0$, $\text{Im}(\eta_3)<0$ in region III. 

Hence, the Stokes curve separating regions I and II switches on the exponential contribution associated with $\eta_2$, the Stokes curve separating regions II and III switches on the exponential contribution associated with $\eta_3$ and the Stokes curve separating regions III and IV switches on the exponential contribution associated with $\eta_1$. To denote the Stokes switching behaviour of these subdominant exponentials, the Stokes curves are labelled by \ding{192}, \ding{193} and \ding{194}. 

In regions I-IV, the presence of the exponential contributions associated with $\eta_j$ are exponentially small since  
$\text{Re}(\eta_j)>0$, and therefore do not affect the dominance of the leading order behaviour in \eqref{type b asym series combination complete}. Hence, the values of $\mathcal{\hat{S}}_j$ may be freely specified in these regions and therefore the asymptotic solution described by \eqref{type b asym series combination complete} contains free parameters hidden beyond-all-orders. 

In Section \ref{S:Stokes structure} we were able to obtain special asymptotic solutions by uniquely specifying the free parameters present in the asymptotic expansion. However, this cannot be done for Type B solutions in the same way because of the possible interaction effects of singularities originating from the different Riemann sheets of $\eta_j$. 

\begin{figure}[h!]
\begin{minipage}{1\linewidth} 
\centering
\scalebox{0.5}{ 
\includegraphics[scale=0.35]{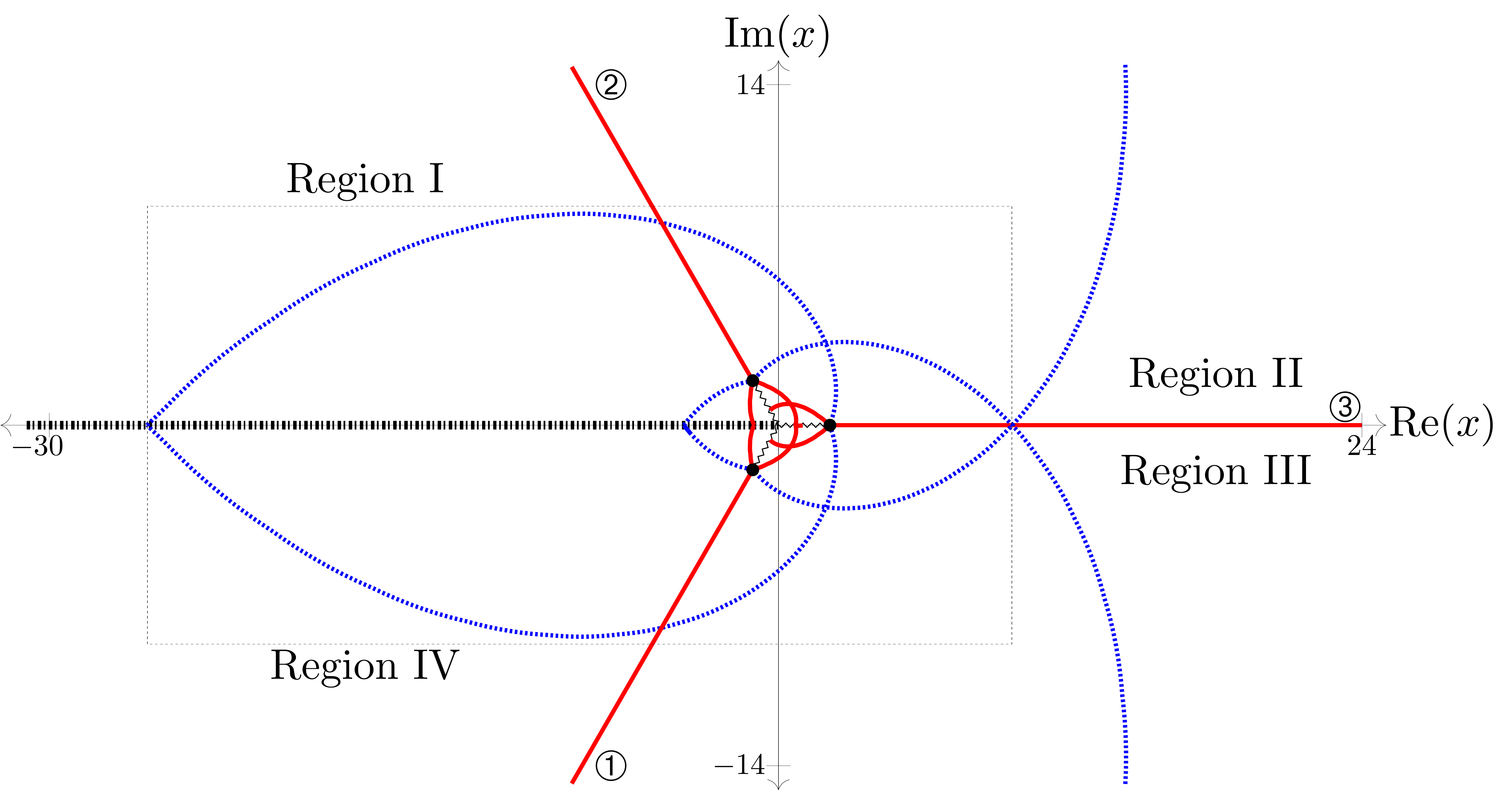}
}
\subcaption{Stokes structure associated with $\eta_j$ in the complex $x$-plane}\label{stokes struc w04 x plane zoom out}
\end{minipage}
\begin{minipage}{0.45\linewidth} 
\centering
\scalebox{0.55}{ 
\includegraphics[scale=0.04]{fig42b}
}
\subcaption{Zoomed out version of Figure \ref{stokes struc w04 x plane zoom out}}\label{stokes struc w04 x plane zoom in}
\end{minipage}
\begin{minipage}{0.45\linewidth} 
\centering
\scalebox{0.55}{ 
\includegraphics[scale=0.04]{fig42c}
}
\subcaption{Regions of validity}\label{stokes struc w04 x regions of valid}
\end{minipage}
\caption{This figure illustrates the Stokes structure of type B solutions of $q$-$\text{P}_\text{I}$ in the original $x$-plane. Under the inverse transformation \eqref{qPI reverse transformation} the Stokes and anti-Stokes curves are described by $q$-spirals in the $x$-plane. The exponential contributions associated with $\eta_1, \eta_2$ and $\eta_3$ are switched across the Stokes curves labelled by \ding{192}, \ding{193} and \ding{194}, respectively as illustrated in Figure \ref{stokes struc w04 x plane zoom in}. In Figure \ref{stokes struc w04 x regions of valid} the regions shaded in blue denote regions in which the asymptotic solution described by \eqref{type b asym series combination complete} is valid. These are the regions in which the exponential contributions present in \eqref{type b asym series combination complete} are exponentially small. The regions shaded in red are regions in which these exponential contributions are exponentially large and hence regions in which the asymptotic behaviour is not described by \eqref{type b asym series combination complete}.}\label{stokes struc w04 x plane TOTAL}
\end{figure}
It may be possible to find special Type B solutions by considering the behaviour of the exponential contributions on the different Riemann sheets of $\eta_j$ and how they interact with those on the principal Riemann sheet. As this is beyond the scope of this study, we restrict our analysis to regions I-IV and therefore only obtain asymptotic solutions described by \eqref{type b asym series combination complete} which contain free parameters hidden beyond-all-orders. 

Following the analysis in Section \ref{S:Stokes structure} we obtain the Stokes structure in the original $x$-plane by applying the inverse transformation given by \eqref{qPI reverse transformation}. As in Section \ref{S:Stokes structure} we demonstrate the Stokes structure for the value of $q=1+0.2i$. Using \textsc{Matlab}, the Stokes structure of Type B solutions is illustrated in Figure \ref{stokes struc w04 x plane TOTAL}.

Figure \ref{stokes struc w04 x plane TOTAL} shows the corresponding regions I-IV in the complex $x$-plane. In this figure, the Stokes and anti-Stokes curves are denoted by the solid red and dashed blue curves respectively. The branch cuts of $\eta_j$ are depicted as the zig-zag curves, which connect the singularities to the origin in the complex $x$-plane. Furthermore, the dot-dashed curve denotes the logarithmic branch cut defined by the reverse transformation \eqref{qPI reverse transformation} for the choice of $q=1+0.2i$. 

Recall that the inverse transformation maps the Stokes and anti-Stokes curves in the complex $s$-plane to $q$-spirals in the complex $x$-plane. In particular, we see in Figure \ref{stokes struc w04 x plane TOTAL} that the Stokes curves of Type B solutions extend to infinity in the complex $x$-plane. This is due to the fact that the Stokes curves in the complex $s$-plane extend to infinity as $\text{Re}(s)\rightarrow \infty$ as shown in Figure \ref{typeB stokes struc s place zoomout}. This was not case for the Stokes curves of Type A solutions in Section \ref{S:Stokes structure}. Instead the Stokes curves of Type A solutions emanate from the singularities and approach the logarithmic branch cuts and enter a different Riemann sheet of the inverse transformation; this is illustrated in Figures \ref{stokes struc w03 x plane}, \ref{stokes struc w01 x plane} and \ref{stokes struc w02 x plane}.

We have therefore determine the regions of validity for Type B solutions of $\text{$q$-P}_\text{I}$, \eqref{qPI rescaled}, in the complex $x$-plane. Type B solutions are described by the asymptotic power series expansion \eqref{type b asym series combination complete} as $\epsilon\rightarrow 0$, and contain free parameters hidden beyond-all-orders. Furthermore, we have also calculated the Stokes behaviour present within these asymptotic solution, \eqref{type b asym series combination complete}, which allowed us to determine regions in which this asymptotic description is valid.

\section{Connection between Type A and Type B solutions and the nonzero and vanishing asymptotic solutions of $q$-Painlev\'{e} I}\label{S: connection to the nonzero asym behaviours of qpI}
In this section we establish a connection between both Type A and B solutions found in this study to the nonzero and vanishing asymptotic solutions of $q$-Painlev\'{e} I respectively. We recall that the nonzero and vanishing asymptotic solutions were first found by Joshi in \cite{Joshi2015}. In particular, equation \eqref{qPainleveI eqn} admits solutions with the following behaviour
\begin{equation}\label{qPI nonzero asym/van shortcut}
w(x)\sim w_{\rm{nv}}(x)= \omega^3 +\mathcal{O}\left(\frac{1}{x}\right), \qquad w(x)\sim w_{\rm{v}}(x)= \frac{1}{x} +\mathcal{O}\left(\frac{1}{x^4}\right),
\end{equation}
as $|x|\rightarrow\infty$, where $w_{\rm{nv}}$ and $w_{\rm{v}}$ are the nonzero and vanishing asymptotic solutions and $\omega^3=1$. 

Recall that the four possible solutions for $W_{0}(s)$ are denoted by $W_{0,j}(s)$, which are defined by \eqref{qPI W01-W02}-\eqref{qPI W03-W04}. In our investigation, we applied the scalings given in \eqref{qPI scalings}, in which the variable $x$ has the behaviour described by \eqref{q difference rescale independent variable} as $\epsilon\rightarrow 0$. The analysis for both Type A and B solutions are valid in the limit $\epsilon\rightarrow 0$, which was shown to be equivalent to the double limit $|q|\rightarrow 1$ and $n\rightarrow\infty$. Under the additional limit, $s\rightarrow+\infty$, we see that the behaviour of $x$ in \eqref{q difference rescale independent variable} approaches infinity. Therefore, the limits $\epsilon\rightarrow 0$ and $s\rightarrow+\infty$ are equivalent to the limit $|x|\rightarrow\infty$. 

We now study the behaviour of $W_{0,j}(s)$ under the additional limit $s\rightarrow+\infty$. By applying the limit $s\rightarrow+\infty$ to the terms appearing in \eqref{qPI leading order component A-C} and \eqref{qPI leading order component D} we find from equations \eqref{qPI W01-W02} and \eqref{qPI W03-W04} that
\begin{equation}\label{type A nonzero 1-4}
\lim\limits_{s\rightarrow+\infty}W_{0,1}=-\frac{1}{2}+i\frac{\sqrt{3}}{2}, \quad \lim\limits_{s\rightarrow+\infty}W_{0,2}=-\frac{1}{2}-i\frac{\sqrt{3}}{2}, \quad \lim\limits_{s\rightarrow+\infty}W_{0,3}=1, \quad \lim\limits_{s\rightarrow+\infty}W_{0,4}=0. 
\end{equation}
The limiting behaviour of Type A solutions under the limit $s\rightarrow+\infty$ are therefore described by cube roots of unity. That is $W_{0,j} \sim \omega^3,$ as $\epsilon\rightarrow 0$ and $s\rightarrow+\infty$ for $j=1,2,3$. However, we also find from \eqref{type A nonzero 1-4} that Type B solutions vanish in the limit $s\rightarrow+\infty$. We therefore find that Type A solutions of \eqref{qPI rescaled} tend to the nonzero asymptotic behaviour solutions, $w_{\rm{nv}}$, found by Joshi \cite{Joshi2015} under the additional limit $s\rightarrow+\infty$. 

In order to calculate the leading order behaviour of $W_{0,4}(s)$ as $s\rightarrow+\infty$, we need to keep terms up to order $\mathcal{O}(e^{-3s})$. Be carefully tracking such terms we find that the behaviour of Type B solutions are given by
\begin{equation}
W_{0,4}\sim\frac{1}{2}\sqrt{1-\frac{4e^{-s}}{3}}-\frac{1}{2}\sqrt{-\left(1-\frac{4e^{-s}}{3}\right)+\frac{2}{1-4e^{-s}/3}}\sim e^{-s}+\mathcal{O}(e^{-3s}), \label{correspondence for type B}
\end{equation}
as $s\rightarrow+\infty$. From \eqref{q difference rescale independent variable} we find that \eqref{correspondence for type B} is equivalent to the behaviour $1/x$ as $|x|\rightarrow\infty$ and hence Type B solutions correspond to the vanishing asymptotic behaviour, $w_{\rm{v}}$, of \eqref{qPainleveI eqn}.

\subsection{Numerical computation for $q$-Painlev\'{e} I}\label{S:numerical aspects of qPI}
In this section we give a numerical example of \eqref{qPI discrete time} with the parameter choice of $q=1+0.2i$ and where $x=x_0q^n$ with $x_0=1$. Given two initial conditions, $w_0$ and $w_1$, a sequence of solutions of \eqref{qPI discrete time} may be obtained by repeated iteration. In general, only a certain choice of initial conditions will give a solution of \eqref{qPI discrete time} which tends to the asymptotic behaviour of interest. We follow the numerical method demonstrated by \cite{Chris2015}, originally based on the works of \cite{Joshi2001} to find appropriate initial conditions which tend to Type A solutions of \eqref{qPI discrete time}.

\begin{figure}[H]
\centering
\begin{minipage}[t]{0.45\linewidth} 
\centering
\scalebox{0.9}{ 
\includegraphics[scale=0.75]{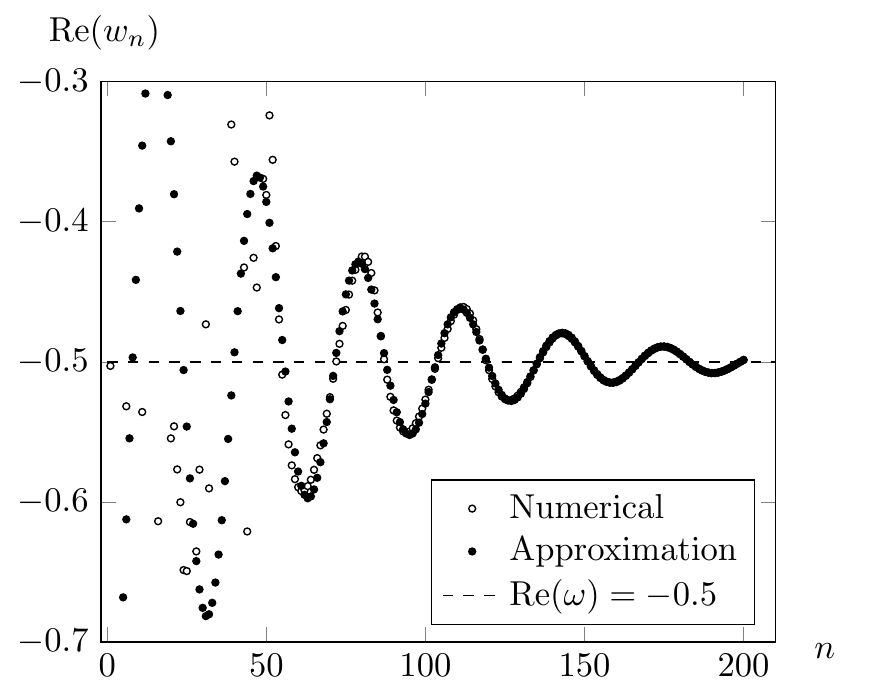}
}
\subcaption{Numerical values of $\text{Re}(w_n)$.}\label{qPIrealnumerics}
\end{minipage}
\begin{minipage}[t]{0.45\linewidth} 
\centering
\scalebox{0.9}{ 
\includegraphics[scale=0.75]{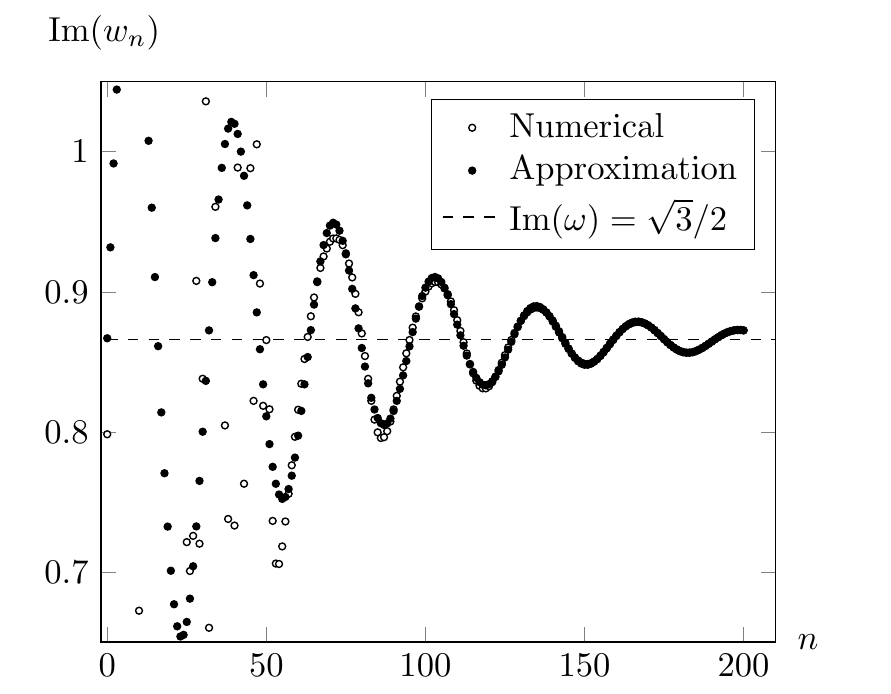}
}
\subcaption{Numerical values of $\text{Im}(w_n)$}\label{qPIimagnumerics}
\end{minipage}
\caption{This figure illustrates the behaviour of the solutions to \eqref{qPI discrete time} with $q=1+0.2i$. The boundary conditions are chosen such that $w_0=0.846885522+i0.798385416$ and $w_1=-0.502881648-i0.650433326$. The values of $w_n$ are represented as blue circles, and the Type A asymptotic solution, $w_n\sim \omega$ as $n\rightarrow\infty$, where $\omega^3=1$, is represented by the black cross marks. From Figures \ref{qPIrealnumerics} and \ref{qPIimagnumerics} we see that the behaviour of the difference equation tends to the asymptotic expression for large $n$.}\label{qPInumerics}
\end{figure}

Figure \ref{qPInumerics} illustrates a comparison between the numerical solution of \eqref{qPI discrete time} with $q=1+0.2i$ and the initial conditions $w_0=0.846885522+i0.798385416$ and $w_1=-0.502881648-i0.650433326$ and the nonzero asymptotic solution, $w_{\text{nv}}$ in \eqref{qPI nonzero asym/van shortcut} with $\omega=(-1+i\sqrt{3})/2$. Figure \ref{qPIrealnumerics} and \ref{qPIimagnumerics} show that real and imaginary part of $w_n$ converges to $-1/2$ and $\sqrt{3}/2$ respectively for large $n$, which is precisely described by the leading order term of $w_{\text{nv}}$ in \eqref{qPI nonzero asym/van shortcut}.

\section{Conclusions}\label{S:Conclusions}
In this study, we extended the exponential asymptotic methods used in \cite{Chris2015,Luu2017} to compute and investigate Stokes behaviour present in the asymptotic solutions of $\text{$q$-P}_\text{I}$ in the double limit $\left|q\right|\rightarrow 1$ and $n\rightarrow\infty$. In order to investigate the solution behaviour of $q$-difference equations we rescaled the variables in the problem such that the double limit $\left|q\right|\rightarrow 1$ and $n\rightarrow\infty$ was equivalent to the limit $\epsilon\rightarrow 0$. We found two types of solutions for $\text{$q$-P}_\text{I}$, which we call Type A and Type B solutions. Type A solutions of $\text{$q$-P}_\text{I}$ are described asymptotically by either \eqref{qPI type A complete asym}, \eqref{qPI type A complete asym W01} or \eqref{qPI type A complete asym W02}, and Type B solutions are described by \eqref{type b asym series combination complete}. The asymptotic descriptions obtained in this analysis are given as a sum of a truncated asymptotic power series and an exponentially subdominant correction term. We then determined the Stokes structure and used this information to deduce the regions of the complex plane in which these asymptotic solutions are valid. 

In Section \ref{S:Exponential Asymptotics} we first considered the asymptotic solutions of \eqref{qPI rescaled} described by Type A solutions. Using exponential asymptotic methods, we determined the form of the subdominant exponential contribution present in the asymptotic solutions, which were found to be defined by one free Stokes switching parameter. From this behaviour, we deduced the associated Stokes structure, illustrated in Figures \ref{qPI singulant behaviours} and \ref{fig remaining type A stokes structure}. By considering the Stokes switching behaviour of these subdominant exponentials, we found that the dominant asymptotic behaviour is described by either \eqref{qPI type A complete asym}, \eqref{qPI type A complete asym W01} or \eqref{qPI type A complete asym W02} in a region in the complex $s$-plane containing the positive real axis. Furthermore, we found that it is possible to select the Stokes parameters so that the exponential contribution is absent in the regions where it would normally be large. Consequently, the regions of validity for the special Type A solutions are larger than the regions of validity for generic Type A solutions as illustrated in Figures \ref{Regions of validity special W03}, \ref{fig type A w01 special regions} and \ref{fig type A w02 special regions}.

In Section \ref{S:Type B Asymptotics}, we considered the equivalent analysis for Type B solutions of \eqref{qPI rescaled}. Compared to Type A solutions, Type B solutions are those which are singular at three points rather than one. Although exponential asymptotic methods may be used again to determine the form of the exponential small contributions present in Type B solutions, we noted that the analysis is near identical except for the determination of the singulant of this problem, $\eta$. The main difference is due to the fact that Type B solutions are singular at three points rather than one, and the remaining analysis was therefore identical as for Type A solutions. Type B asymptotic solutions are given as a sum of a truncated asymptotic power series and three exponentially subdominant correction terms as described in \eqref{type b asym series combination complete}. The Stokes structure for Type B solutions, illustrated in Figure \ref{stokes struc w04 s plane}, is significantly more complicated than the Stokes structure of Type A solutions. In order to describe the Stokes switching behaviour in the domain $\mathcal{D}_0$ we must understand how the asymptotic solution \eqref{type b asym series combination complete} interacts with solutions on different Riemann sheets. However, we restrict ourselves to regions I-IV as these regions are free of possible interaction effects with singularities originating from the different Riemann sheets. Furthermore, all three exponential contributions present in Type B solutions are exponentially subdominant in regions I-IV and therefore represent the regions of validity of \eqref{type b asym series combination complete}. Consequently, the asymptotic solutions described by \eqref{type b asym series combination complete} contain one free parameter defined by the Stokes multiplier.  

By reversing the rescaling transformations we were then able to obtain the corresponding Stokes structure in the complex $x$-plane and found that the Stokes and anti-Stokes curves are described by $q$-spirals in the complex $x$-plane. As a result, the regions of validity are no longer described by traditional sectors bounded by rays but sectorial regions bounded by arcs of spirals. Consequently, this methodology is applicable to other $q$-Painlev\'{e} equations or more generally nonlinear $q$-difference equations in order to obtain asymptotic solutions which display Stokes phenomena.

Finally, Section \ref{S: connection to the nonzero asym behaviours of qpI} demonstrated that Type A and B solutions are related to the nonzero asymptotic and vanishing asymptotic solutions found by Joshi \cite{Joshi2015}. If the additional limit $s\rightarrow+\infty$ is also taken, then we find that Type A solutions correspond to the nonzero asymptotic solutions of $\text{$q$-P}_\text{I}$ while Type B solutions correspond to the quicksilver solutions of $\text{$q$-P}_\text{I}$ \cite{Joshi2015}.

\begin{appendices}

\section{Calculating the late-order terms near the singularity}\label{S:Appendix Calculate late order}
In order to determine the value of $\gamma_1$ in \eqref{even late order terms} we calculate the local behaviour of $\chi_3$ and $U_3$. Using \eqref{local qPI leading order behaviour} in equations \eqref{qPI singulant expressions} and \eqref{qPI EXACT PREFACTOR} we can show that
\begin{align}
\chi_3(s)\sim \frac{4i\sqrt{6\sqrt{2}}}{5}(s-s_{0,3})^{5/4}, \qquad
U_3(s)\sim \frac{\Lambda}{\sqrt{i\sqrt{6\sqrt{2}}}(s-s_{0,3})^{1/8}}, \label{local even prefactor behaviour}
\end{align}
as $s\rightarrow s_{0,3}$. Furthermore, by calculating the expression for $W_1$ (using equation \eqref{even recurrence relation}) it can be shown that $W_1(s)=\mathcal{O}\left((s-s_{0,3})^{-1/2}\right)$ and from \eqref{integral expression in prefactor} that 
\begin{equation}\label{local prefactor exponential behaviour}
e^{\pm G(s)}= 1\pm A_1(s-s_{0,3})^{1/4}+\mathcal{O}\left((s-s_{0,3})^{1/2}\right),
\end{equation}
as $s\rightarrow s_{0,3}$ for some constant $A_1$. By substituting \eqref{local even prefactor behaviour} into the expression for $W_r$ as given by \eqref{late order terms with sing and prefac} we find that the local behaviour of of the late-order terms near the singularity is given by 
\begin{align}
W_r(s)\sim& \left(\frac{5}{4}\right)^{r+\gamma_1}\frac{\Gamma(r+\gamma_1)}{\left(i\sqrt{6\sqrt{2}}\right)^{r+\gamma_1+1/2}(s-s_{0,3})^{5(r+\gamma_1)/4+1/8}}\left(\Lambda e^{-G}+\frac{\tilde{\Lambda}e^{G}}{(-1)^{r+\gamma_1}}\right), \label{even late order terms near sing}
\end{align}
as $s\rightarrow s_{0,3}$. From \eqref{even recurrence relation}, the dominant behaviour of $W_r$ is due to the term $W_{r-2}''/(4W_0^3-1)$. Hence, if $W_{r-2}$ has a singularity of strength $\nu$, then $W_r$ has one of strength $\nu+5/2$. In particular, since we know that $W_0$ is singular at $s_{0,3}$ of strength $-1/2$ then $W_{2r}$ will also be singular at $s_{0,3}$ but with strength $5r/2-1/2$. Moreover, since $W_1$ is also singular at $s_{0,3}$ of strength $1/2$, then $W_{2r+1}$ has one of strength $5r/2+1/2$. 

We first observe from \eqref{local prefactor exponential behaviour} that 
\begin{align}\label{local prefactor exponential behaviour odd and even}
\left(\Lambda e^{-G}+\frac{\tilde{\Lambda}e^{G}}{(-1)^{r+\gamma_1}}\right)\sim\begin{cases*}
\Lambda+i\tilde{\Lambda}+A_1(-\Lambda+i\tilde{\Lambda})(s-s_{0,3})^{1/4}+\mathcal{O}((s-s_{0,3})^{1/2}), \quad \text{for even $r$,} \\
\Lambda-i\tilde{\Lambda}+A_1(-\Lambda-i\tilde{\Lambda})(s-s_{0,3})^{1/4}+\mathcal{O}((s-s_{0,3})^{1/2}), \quad \text{for odd $r$,}
\end{cases*}
\end{align}
as $s\rightarrow s_{0,3}$. Thus, in order for the singularity behaviour of \eqref{even late order terms near sing} to be consistent with that of $W_{2r}$, we require that $5r/2-1/2=5(2r+\gamma_1)/4+1/8$ under the condition $\Lambda+i\tilde{\Lambda}\neq0$.  Hence, we deduce that $\gamma_1=-1/2$. Using the same argument, it can be shown that in order for $W_{2r+1}$ to have the correct singular behaviour in the limit $s\rightarrow s_{0,3}$ we must impose the condition $\Lambda-i\tilde{\Lambda}=0$, which gives 
\begin{equation}\label{upsilson condition}
\tilde{\Lambda}=-i\Lambda. 
\end{equation} 

\section{Calculating the prefactor constants via the inner problem}\label{S:Appendix calculate prefactors}
The expression for the late-order terms given by \eqref{even late order terms} contains a constant $\Lambda_3$, which is yet to be determined. In order to determine the value of $\Lambda_3$ we perform an inner analysis of \eqref{qPI rescaled} near the singularity, $s_{0,3}$ and determine the inner expansion of the inner solution. We then use the method of matched asymptotics to match the outer expansion to the inner expansion following Van Dyke's matching principle \cite{Hinch1991}.

In view of the leading order behaviour given by \eqref{local qPI leading order behaviour}, we study the inner solution by applying the scalings
\begin{equation}\label{qPI inner rescalings}
s= s_{0,3}+\epsilon^{4/5}\zeta, \qquad W(s)=a_3+\epsilon^{2/5}\psi_1(\zeta)+\epsilon^{3/5}\psi_2(\zeta),
\end{equation}
where $\zeta$ is the inner variable, $\psi_1$ and $\psi_2$ are the first two terms of the inner solution, $\psi$. In particular, we recall that 
\begin{equation}\label{qPI inner parameter reminders}
s_{0,3}=\frac{1}{3}\log\left(\frac{256}{27}\right), \quad a_3=\left(\frac{1}{4}\right)^{1/3}, \quad b_3=\left(\frac{1}{8\sqrt{2}}\right)^{1/3}.
\end{equation}
Substituting \eqref{qPI inner rescalings} into \eqref{qPI rescaled} we obtain 
\begin{align}
(a_3^4-a_3+e^{-s_{0,3}})+\epsilon^{2/5}(4a_3^3-1)\psi_1+&\epsilon^{3/5}(4a_3^3-1)\psi_2+\epsilon^{4/5}(6a_3^2\psi_1^2-e^{-s_{0,3}}\zeta+a_3^3\psi_1'')\nonumber \\
+&\epsilon\left(\frac{s_{0,3}}{2}e^{-s_{0,3}}+12a_3^2\psi_1\psi_2+a_3^3\psi_2''\right)+\mathcal{O}(\epsilon^{6/5})=0, \label{qPI inner matching orders}
\end{align}
as $\epsilon\rightarrow 0$, and where the prime denotes derivatives with respect to $\zeta$. Using the values of $a_3, b_3$ and $s_{0,3}$ given in \eqref{qPI inner parameter reminders} we find that the coefficients of $\epsilon^0, \epsilon^{2/5}$ and $\epsilon^{3/5}$ are identically zero. Therefore, the leading order equation of the inner solution is given by
\begin{equation}\label{even qPI inner solution equation}
\psi_1^2-b_3^2\zeta+\frac{a_3}{6}\frac{d^2\psi_1}{d\zeta^2}=0,
\end{equation}
as $\epsilon\rightarrow 0$. From equation \eqref{qPI inner matching orders} we see that the term $\psi_2$ does not appear in the leading order equation in the limit $\epsilon\rightarrow \infty$. This reinforces the fact that the odd terms are indeed negligible in the limit $\epsilon\rightarrow 0$ as the term $\psi_2$ corresponds to the first odd coefficient term in the outer problem (far field expansion), \eqref{qPI asymptotic power series}. 

To study the inner solution, we analyze \eqref{even qPI inner solution equation} in the limit $\left|\zeta\right|\rightarrow \infty$. Using the method of dominant balance, equation \eqref{even qPI inner solution equation} has a solution described by $\psi_1 \sim b_3\sqrt{\zeta}$ as $\left|\zeta\right|\rightarrow \infty$. For algebraic convenience, we rescale the inner solution by setting
\begin{equation}\label{qPI inner solution series expansion}
\psi_1(\zeta) \sim b_3\sqrt{\zeta}\sum_{r=0}^{\infty}\frac{E_r}{\zeta^{5r/2}}, 
\end{equation}
with $E_0=1$. We then substitute \eqref{qPI inner solution series expansion} into \eqref{even qPI inner solution equation} and match terms of $\mathcal{O}(\zeta)$ in the limit $\left|\zeta\right|\rightarrow \infty$. Doing this, we obtain the following nonlinear recurrence relation
\begin{equation}\label{even qPI inner solution coefficients}
E_r = -\frac{1}{2b_3^2}\left(\frac{a_3b_3}{24}(5r-4)(5r-6)E_{r-1}+b_3^2\sum_{k=1}^{r-1}E_{r-k}E_k\right),
\end{equation}
for $r\geq 1$. We can express \eqref{qPI inner solution series expansion} in terms of the outer variables by reversing the scalings given in \eqref{qPI inner rescalings}. Doing this, we find that the inner expansion of the outer solution is given by
\begin{equation}\label{qPI inner solution series expansion FINAL}
W(s)\sim a_3+b_3\sum_{r=0}^{\infty}\frac{\epsilon^{2r}E_r}{(s-s_{0,3})^{(5r-1)/2}},
\end{equation}
as $\epsilon\rightarrow 0$. Recall that the outer expansion is given by the expression \eqref{qPI asymptotic power series} as $\epsilon\rightarrow0$, and where the behaviour of coefficients are given by \eqref{explicit LOT}. By matching the expansions \eqref{qPI asymptotic power series} and \eqref{qPI inner solution series expansion FINAL} it follows that 
\begin{equation}\label{even qPI late order terms CONSTANT}
\Lambda_3 = \lim\limits_{r\rightarrow\infty} b_3 E_r \frac{\sqrt{i\sqrt{6\sqrt{2}}}}{\Gamma(2r-1/2)}\left(\frac{4i\sqrt{6\sqrt{2}}}{5}\right)^{2r-\tfrac{1}{2}}.
\end{equation}

\begin{figure}[H]
\centering
\includegraphics[scale=0.8]{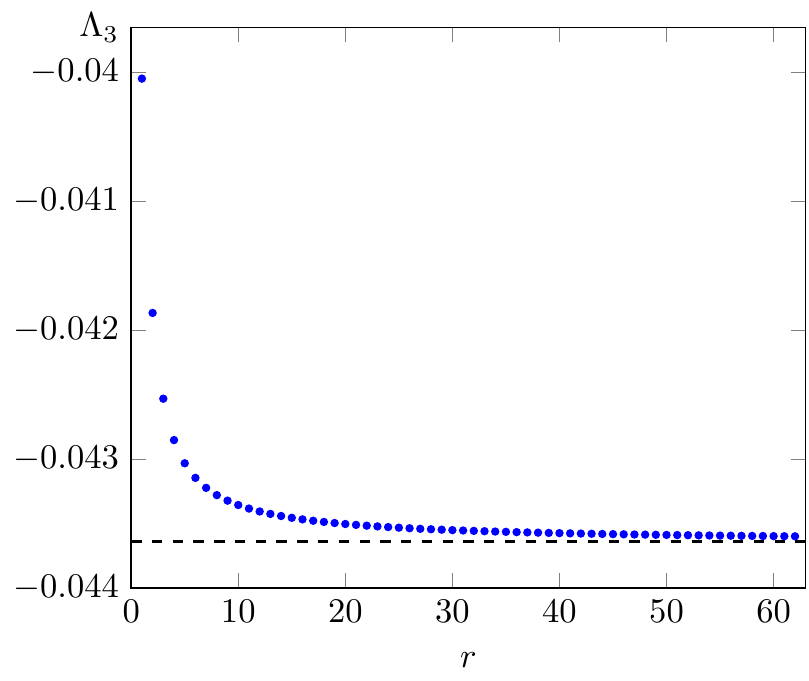}
\caption{This figure illustrates the approximation for $\Lambda$ appearing in the even late-order terms \eqref{even late order terms near sing}. As $r$ increases, the approximation for $\Lambda$ tends to the limiting value of $-0.04364$, which is denoted by the black dashed line.}\label{chapt 4 fig 1}
\end{figure}

We then compute the first 1000 $E_r$ terms using the recurrence relation \eqref{even qPI inner solution coefficients}. Then, by using the formula \eqref{even qPI late order terms CONSTANT} we find numerically that the approximate value of $\Lambda_3$ is 
\begin{equation}\label{qPI lambda value}
\Lambda_3\approx -0.04364,
\end{equation}
to four significant figures. The approximate value for $\Lambda_3$ is shown in Figure \ref{chapt 4 fig 1}.

\section{Stokes smoothing}\label{S:Appendix Stokes smoothing}
To apply the exponential asymptotic method, we optimally-truncate the asymptotic series \eqref{qPI asymptotic power series}. One particular way to calculate the optimal truncation point is to consider where the terms in the asymptotic series is at its smallest \cite{Boyd1999}. This heuristic is equivalent to the finding $N$ such that   
\begin{equation*}
\bigg|\frac{\epsilon^{2N+2}W_{2N+2}}{\epsilon^{2N} W_{2N}}\bigg|\sim 1,
\end{equation*}
in the limit $N\rightarrow\infty$. By using the late-order form ansatz described by \eqref{even late order terms} we find that $N\sim |\chi_3|/(2\epsilon)$ as $\epsilon\rightarrow 0$ (which is equivalent to the limit $N\rightarrow\infty$). As this quantity may not necessarily be integer valued, we therefore choose $\kappa\in[0,1)$ such that 
\begin{equation}\label{optimal truncation point}
N_{\text{opt}} \sim \frac{1}{2}\left(\frac{\left|\chi_3\right|}{\epsilon}+\kappa\right),
\end{equation} 
is integer valued.

By substituting the truncated series, \eqref{qPI optimally truncated series}, into the governing equation \eqref{qPI rescaled} we obtain 
\begin{equation}\label{governing qPI remainder equation part 1}
\overline{T}T^2\underline{T}+T^2\underline{T}\overline{R}+2\overline{T}T\underline{T}R+\overline{T}T^2\underline{R}+\cdots=T+R-\frac{1}{(1+\epsilon)^{s/\epsilon}},
\end{equation}
where the terms neglected are quadratic in $R_N$. We can use the recurrence relation \eqref{even recurrence relation} to cancel terms of size $o(\epsilon^{2N-1})$ in \eqref{governing qPI remainder equation part 1}. In particular, equation \eqref{governing qPI remainder equation part 1} can be rewritten, after rearrangement, as
\begin{equation}\label{governing qPI remainder equation appendix}
T^2\underline{T}\overline{R}+2\overline{T}T\underline{T}R+\overline{T}T^2\underline{R}-R \sim \epsilon^{2N}(4W_{0,3}^3-1)W_{2N},
\end{equation}
as $\epsilon\rightarrow 0$ and where the terms neglected are of order $\mathcal{O}(\epsilon^{2N+1}W_{2N+1})$ and terms quadratic in $R_N$. Terms of these size are negligible compared to the terms kept in \eqref{governing qPI remainder equation appendix} as $\epsilon\rightarrow 0$.

Away from the Stokes curve the inhomogeneous terms of equation \eqref{governing qPI remainder equation appendix} are negligible as $\epsilon\rightarrow 0$. We may therefore apply a WKB analysis to the homogeneous version of \eqref{governing qPI remainder equation appendix} by setting $R_{N,\text{hom}}(s)=\alpha(s)e^{\beta(s)/\epsilon}$ as $\epsilon\rightarrow 0$. Using the WKB method, we find that the leading order equation gives the solution $\beta=-\chi_3$. Continuing to the next order involves matching terms of $\mathcal{O}(\epsilon R_N)$. By collecting terms of this size in the homogeneous version of \eqref{governing qPI remainder equation appendix} we obtain the equation 
\begin{equation}\label{qPI WKB prefactor equation}
-2W_{0,3}^3\sinh(\chi_3')\frac{\alpha'}{\alpha}-W_{0,3}^3\chi_3''\cosh(\chi_3')+2W_{0,3}^2W_{0,3}'\sinh(\chi_3')+3\frac{W_1}{W_{0,3}}=0,
\end{equation}
where we have substituted the fact that $\beta=-\chi_3$. Comparing equations \eqref{qPI WKB prefactor equation} and \eqref{qPI PREFACTOR DE} show that $\alpha$ satisfies to same differential equation as the prefactor $U_3$ and hence $\alpha \propto U_3$. Hence, away from the Stokes curves the solution of \eqref{governing qPI remainder equation appendix} is given by 
\begin{equation*}\label{qPI remainder away from Stokes}
R_{N,\text{hom}}(s)\sim U_3(s)e^{-\chi_3(s)/\epsilon},
\end{equation*}
as $\epsilon\rightarrow 0$. 

In order to determine the Stokes switching behaviour associated with the optimally-truncated error, we set 
\begin{equation}\label{qPI Stokes Switching WKB}
R(s) = \mathcal{S}_3(s)R_{N,\text{hom}}(s),
\end{equation}
where $\mathcal{S}_3(s)$ is the Stokes multiplier. We substitute \eqref{qPI Stokes Switching WKB} into \eqref{governing qPI remainder equation appendix} and use equations \eqref{qPI SINGULANT DE} and \eqref{qPI PREFACTOR DE} to cancel terms. Doing this we find that the Stokes multiplier satisfies
\begin{align}
\frac{d\mathcal{S}_3}{ds}&\sim\frac{\epsilon^{2N-1}(1-4W_{0,3}^3)W_{2N}}{2W_{0,3}^3\sinh(\chi_3')\alpha(s)}e^{\chi_3/\epsilon}\sim \epsilon^{2N-1}\sqrt{1-4W_{0,3}^3}\frac{\Gamma(2N+\gamma_1)}{\chi_3^{2N+\gamma_1}}e^{\chi_3/\epsilon}, \label{qPI Stokes Multiplier Equation}
\end{align}
as $\epsilon\rightarrow 0$.

Noting the form of $N$, we introduce polar coordinates by setting $\chi_3=\rho e^{i\theta}$ where the fast and slow variables are $\theta$ and $\rho$, respectively. This transformation tells us that
\begin{equation}\label{change of variables}
\frac{d}{ds}=-\frac{i\chi_3'e^{-i\theta}}{\rho}\frac{d}{d\theta},
\end{equation}
and \eqref{optimal truncation point} becomes $N_{\rm{opt}}=(\rho/\epsilon+\kappa)/2$. Under this change of variables equation \eqref{qPI Stokes Multiplier Equation} becomes
\begin{equation}\label{qPI Stokes multiplier asymptotic DE}
\frac{d\mathcal{S}_3}{d\theta}\sim \frac{i\sqrt{2\pi\rho}\sqrt{1-4W_{0,3}^3}}{\chi_3'}\exp\left(\frac{\rho}{\epsilon}(e^{i\theta}-1-i\theta)-i\theta(\kappa-3/2)\right),
\end{equation}
as $\epsilon\rightarrow 0$, where we have used Stirling's approximation \cite{Abram2012} of the Gamma function. For simplicity, we will let $H(s(\theta);\rho)=\sqrt{1-4W_{0,3}^3}/\chi_3'$. From \eqref{qPI Stokes multiplier asymptotic DE} we find that the right hand side is exponentially small everywhere except in the neighbourhood of $\theta=0$, which is exactly where the Stokes curve lies (where $\chi_3$ is purely real and positive). We now rescale about the neighbourhood of the Stokes curve in order to study the switching behaviour of $\mathcal{S}_3$ by setting $\theta = \sqrt{\epsilon}\hat{\theta}$. Note that under this scaling, $H(s(\theta);\rho)\sim H(|\chi_3|)$ as $\epsilon\rightarrow 0$, which is therefore independent of $\theta$ to leading order. Applying the scaling $\theta=\sqrt{\epsilon}\hat{\theta}$ to \eqref{qPI Stokes multiplier asymptotic DE} gives
\begin{equation}\label{qPI Stokes multiplier asymptotic DE apply scaling}
\frac{1}{\sqrt{\epsilon}}\frac{d\mathcal{S}_3}{d\hat{\theta}}\sim i\sqrt{2\pi\rho}H(|\chi_3|)\exp\bigg(-\frac{|\chi_3| \hat{\theta}^2}{2}\bigg),
\end{equation}
as $\epsilon\rightarrow 0$. Integrating \eqref{qPI Stokes multiplier asymptotic DE apply scaling} we find that
\begin{align}
\mathcal{S}_3&\sim i\sqrt{2\pi\epsilon}H(\left|\chi_3\right|)\int_{}^{\left|\chi_3\right|\hat{\theta}}e^{-x^2/2}dx=i\pi\sqrt{\epsilon} H(\left|\chi_3\right|)\left(\text{erf}\left(\theta\sqrt{\frac{\left|\chi_3\right|}{2\epsilon}}\right)+C_3\right), \label{qPI Stokes multiplier integral}
\end{align}
where $C_3$ is an arbitrary constant. Thus, as Stokes curves are crossed, the Stokes multiplier changes in value by
\begin{align*}\label{qPI Stokes Multiplier JUMP appendix}
\Delta\mathcal{S}_3\sim& 2i\pi\sqrt{\epsilon} H(\left|\chi_3\right|),
\end{align*}
and therefore
\begin{equation*}\label{qPI remainder JUMP}
\Delta R_N \sim 2i\pi\sqrt{\epsilon} H(\left|\chi_3\right|)U_3(s)e^{-\chi_3(s)/\epsilon},
\end{equation*}
as $\epsilon\rightarrow 0$.

\end{appendices}

\bibliographystyle{plain}
\bibliography{Main_bibliography}

\end{document}